\documentclass[12pt]{article}
\usepackage{epsf,amsmath}
\usepackage{amssymb}
\usepackage{indentfirst}
\usepackage{array,tabularx}
\makeatletter \@addtoreset{footnote}{page} \makeatother

\begin{document}

\title{\large \bf Analysis of Light Curves of Eclipsing Systems
with Exoplanets: HD 189733}

\author{\large\bf M. K. Abubekerov, N. Yu. Gostev, and A. M. Cherepashchuk \\
\normalsize\it Sternberg Astronomical Institute, Moscow State
University \\ marat@sai.msu.ru, ngostev@mail.ru, cher@sai.msu.ru}

\date{\begin{minipage}{15.5cm} \small
High-accuracy multicolor light curves of the binary system HD
189733, which contains an exoplanet, are analyzed. We have
determined the radii of the star and the planet in the binary
system, as well as the orbital inclination. The limb-darkening
coefficients of the stellar disk were obtained in 10 filters in
the wavelength interval $\lambda\lambda=5500-10500 \AA$. The
uncertainties of the fitted parameters were estimated using the
differential-correction and confidence-area methods. The
wavelength dependence of the limb-darkening coefficients is
compared to the corresponding theoretical function for a model
thin stellar atmosphere. We confirm the wavelength dependence of
the exoplanet's radius found by Pont et al. \cite{Pont2008} (at
the $1\sigma$ level). The exoplanet radius increases with
decreasing wavelength, which seems to argue for the presence of an
atmosphere around the planet.
\end{minipage}}\maketitle \rm

\section*{\normalsize INTRODUCTION }

The results of fitting high-accuracy multicolor light curves for
the star HD 209458, which is eclipsed by an exoplanet, were
provided in \cite{Abubekerov2010}. There was a significant
deviation between the observed and theoretical limb-darkening
coefficients, even though the most conservative values were used
to estimate the "external" uncertainties in the model parameters
\cite{Claret2004, Claret2009}. This confirms the discrepancy
between observations and the theory of thin stellar atmospheres
found in \cite{Southworth2008}. It is important to analyze limb
darkening across other stellar disks that are eclipsed by their
exoplanets to further study this effect. Pont et al.
\cite{Pont2007, Pont2008} obtained high-accuracy multicolor light
curves of HD 189733, which is eclipsed by an exoplanet, and
examined the spotted disk of the star. Pont et al. \cite{Pont2007,
Pont2008} indicated that the exoplanet could have an atmosphere.
The purpose of the current paper is to study the limb darkening of
the star in the eclipsing system HD 189733 as a function of
wavelength in detail using the data obtained in \cite{Pont2007}.
We analyze in detail the effect of uncertainties in the parameters
determined using different methods to fit the eclipsing light
curves, as in \cite{Abubekerov2010}.

\section*{\normalsize FITTING METHOD}

The method used to fit transit light curves observed in eclipsing
binary systems containing exoplanets is described in detail in
\cite{Abubekerov2010}, and we will only consider the main concepts
of the method here. We used a model consisting of two spherical
stars in circular orbits, without any reflection effect or
ellipsoidal effect. There are no spots on the surfaces of the
stars. If the masses of a star and planet are $m_s = 0.825
M_{\odot}$ and $m_p = 1.15 M_{Jup}$ \cite{Bouchy2005}, the mean
relative radius of the Roche lobe for the planet calculated using
the formula of Eggleton \cite{Eggleton1983} is $R_R/a = 0.0531$,
where $a$ is the radius of the relative orbit of the system.

Since the relative radius of the planet is $r_p = R_p / a \simeq
0.0175$ (see below), the planet fills its Roche lobe only to the
degree $\mu_p \simeq 0.33$, which is considerably less than 0.5.
Therefore, a spherical approximation is quite satisfactory for the
planet (neglecting flattening of the planet due to its axial
rotation). The same can be said for the optical star.

The amplitude of brightness changes due to the reflection effect
in the optical should be less than $10^{-5}$ magnitude at eclipse
phases, which is negligible \cite{Abubekerov2010}. A weak
wavelength dependance of the radius of the exoplanet in HD 189733
was suggested in \cite{Pont2008}, which could be indicative that
the planet has an atmosphere. The refraction of the stellar light
in the exoplanet's atmosphere can distort the light curve when the
star is eclipsed by the exoplanet. The effects of light refraction
in eclipsing binary systems were calculated by Kudzei
\cite{Kudzei1985}. These effects should result in small humps
(brightenings) in the light curve before and after the eclipse, as
well as in the middle of the eclipse. Since there are no such
humps in the light curves of HD 189733, refraction effects in
these light curves can be assumed to be negligible.

When the star is eclipsed by its exoplanet, the light curve could
be affected by gravitational microlensing. Such effects were
studied in \cite{Kasuya2011}. Microlensing was shown to be
significant only for the stars with exoplanets whose orbital sizes
exceed 10 AU (with corresponding orbital periods of $P>10$ yr).
Therefore, our model with two spherical components is quite
applicable to the light curves of HD 189733.

We calculated the light curve using linear and quadratic
limb-darkening laws to describe the brightness distribution across
the stellar disk:

\begin{equation}
I(\rho)=I_0 \left(1-x+x\sqrt{1-\frac{\rho^2}{r_s^2}} \right),
\label{I1}
\end{equation}
\noindent
\begin{equation}
I(\rho) = I_0 \left(1-x
\left(1-\sqrt{1-\frac{\rho^2}{r_s^2}}\right) - y\left(
1-\sqrt{1-\frac{\rho^2}{r_s^2}} \right)^{\!\!\!2} \right)\,.
\label{Ixy}
\end{equation}

Here, $\rho$ is the polar distance from the center of the stellar
disk, $I_0$ the brightness at the center of the disk, $r_s$ the
radius of the stellar disk in radii of the relative orbit, and x
the limb-darkening coefficient. The brightness at the center of
the planetary disk, and accordingly the brightness at any point of
this disk, are assumed to be zero. The planet eclipses the star
when the orbital phase is $\theta=\pi$. The orbit is taken to be
circular, and the radius of the relative orbit $a$ to be unity.
The model does not include "third light." The radius of the planet
in relative orbital radii is denoted $r_p$. The desired parameters
of our model are the radius of the star $r_s$, the radius of the
planet $r_p$, the inclination of the orbit to the plane of the sky
$i$, and the limb-darkening coefficient $x$; if the limb darkening
is described by the quadratic function, the parameters also
include the coefficient $y$.

The total light of the system is taken to be unity, and the
observed brightness values to be normally distrubuted. The rms
deviations of the observed brightnesses $\sigma$ are assumed to be
known.

To search for the optimal (central) values of the model parameters
and generate their confidence intervals, we used a residual equal
to the sum of the squared differences between the observed and
theoretical brightnesses (as functions of the desired parameters),
divided by the dispersions of the observed brightnesses (the
squares of the rms deviations $\sigma$). If amodel fits the
observational data, this residual is distributed according to a
$\chi^2_M$ law for the exact parameter values, where $M$ is the
number of observational points. This residual (which is a convex
function of the parameters) achieves its minimum for certain
parameters values $r^c_s, r^c_p, i^c, x^c, y^c$, referred to as
the central (or optimal) parameters. When the model fits the
observational data, the distribution of the central parameter
values can be considered to be normal in some vicinity of their
mean values, if we neglect non-linearity of the
parameter-dependent brightness in this neighbourhood. This is
justified when $\sigma$ is small and the number of observational
points $M$ is large.

The central parameter values are random values distributed
normally in this approximation. They can be used as a statistic in
the differential-correction method or Monte-Carlo method (which,
in a sense, is equivalent to the differential-correctionmethod
\cite{Abubekerov2008, Abubekerov2009}). The error intervals for
the parameters are chosen to be centered on the optimal values of
the corresponding parameters; their half-widths are equal to the
mean square estimates of the rms deviations of the current
parameters $\sigma_{est}$ from their central values, multiplied by
a coefficient corresponding to the chosen confidence level
$\gamma$ ($1\sigma, 2\sigma, 3 \sigma$, etc.)
\cite{Abubekerov2009}. In this case, the model is considered to be
perfectly valid; if we have more than one parameter, each
individual parameter is characterized by its individual confidence
interval, which encompasses the true value of that parameter with
a given probability $\gamma$ (independent of whether or not the
true values of the other fitted parameters fall into their
confidence intervals). We also assume that we can neglect
non-linearity of the function describing the dependence of the
brightness on the parameters within approximately one rms
deviation around the central parameters. Numerical experiments
carried out for eclipsing systems show that the probability for
all the true parameter values to simultaneously fall into their
confidence intervals is a factor of $1.2 \div 1.5$ lower than the
given confidence level $\gamma$ \cite{Abubekerov2008}. Since in
the differential-correction method (or Monte-Carlo method) the
obtained central parameter values are distributed statistically
according to a normal law with the strict a priori assumption that
the model is perfectly valid, and information on the dispersion of
the observed brightnesses is not used here (the confidence
intervals are constructed using the mean square estimates of the
central parameters $\sigma_{est}^2$ rather than the dispersions of
their central values), the uncertainties of the parameters in the
differentialcorrection method (or Monte-Carlo method) are
"internal" uncertainties. As a rule, these are substantially
underestimated. Popper \cite{Popper1984} indicated that this
underestimation can reach a factor of three to five for eclipsing
systems.

In view of the above, we also used the confidencearea method to
estimate the parameter uncertainties \cite{Cherepashchuk1993}. The
advantage of this method is that the obtained confidence intervals
for the parameters ensure that the exact values of all the
parameters simultaneous fall into the corresponding confidence
area (in a multidimensional parametric space) with the given
probability $\gamma$. The probability that the exact value of an
individual parameter falls into the corresponding confidence
interval is greater than the given probability $\gamma$. In
addition, the probability for the exact solution (a combination of
all fitted parameters) to be covered simultaneously by all the
confidence intervals is also greater than $\gamma$, since the
volume of the parallelepiped in which the confidence area $D$ is
inscribed exceeds the volume of this area.

Therefore, the confidence-area method yields more conservative
"external"\, estimates of the parameter uncertainties, which are a
factor of a few greater than the error estimates obtained using
the differentialcorrection method orMonte-Carlo method.

We will use either a $\chi_P^2$ or a $\chi_M^2$ distribution to
find the confidence area $D$ \cite{Abubekerov2008,
Abubekerov2009}. We will use the difference between the residual
distributed according to the $\chi^2_M$ law ($M$ is the number of
observational points) and the minimal value achieved with the
central parameter values as the $\chi^2_P$ statistic, where $P$ is
the number of fitted parameters. If we neglect the non-linearity
in the dependence of the brightness on the parameters (which is
justified when there are many observational points $M$), these
differences can be considered to be distributed according to the
$\chi^2_P$ law for the exact parameter values, where $P=4$ with
linear limb darkening and $P=5$ with quadratic limb darkening.

When using statistics distributed according to $\chi^2_M$ or
$\chi^2_P$, the confidence area $D$ is taken to be a
multidimensional set in the parameter space, for which the
specified statistics are less than the quantile corresponding to
the given confidence level $\gamma$. The probability for $D$ to
encompass the combined exact parameter values is then $\gamma$.
The confidence intervals (uncertainties) for the parameters
corresponding to a given confidence level $\gamma$ are the
projections of the sides of the $P$-dimensional parallelepiped (in
which the confidence area $D$ is inscribed) onto the coordinate
axes in the parameter space \cite{Abubekerov2008}. As was noted
above, this method for constructing the confidence intervals
ensures that they will simultaneously encompass the exact values
of all the parameters with the given probability $\gamma$ (or
higher). Therefore, when specifying the confidence intervals
(uncertainties) for the parameters, we should bear in mind that
the adopted $\gamma$ (for the differential-correction or Monte-
Carlo method) is equal to the probability for the exact value of
each fixed parameter to be encompassed by the corresponding
confidence interval; at the same time, the probability for the
exact solution to be encompassed simultaneously in the all
confidence intervals is less than $\gamma$. When we use the
confidence area method, the probability for the exact value of a
fixed parameter to be included in each individual confidence
interval is greater than the adopted $\gamma$, whereas the
probability that the confidence area $D$ encompasses the combined
exact values of all the parameters is $\gamma$. Therefore, the
adopted confidence level $\gamma$ specifies each of the individual
confidence levels in the differential-correction method and
Monte-Carlo method, and is related to the total confidence area
$D$ in the confidence-area method.

Similar to the differential-correction method, using the
$\chi^2_P$ statistic assumes that a model is perfectly valid, and
that the obtained asymptotic confidence area $D$ never degenerates
into an empty set. The confidence area $D$ does not depend on the
sample size, and the probability $\gamma$ can be specified for an
area $D$ that includes the exact solution (this probability is
referred to as the confidence level). Since it is strictly assumed
that the model is perfectly valid (the residuals are minimized
over the parameters), the parameter uncertainties obtained using
the $\chi^2_P$ statistic can also be considered to be "internal,"
though these are more conservative than the uncertainties obtained
using the differential-correction or Monte-Carlo method. We must
bear in mind, however, that the dispersions of the observational
data are used to construct the $\chi^2_P$ statistic; this is a
significant difference between the use of the $\chi^2_P$
distribution and the differential-correction method. This
difference can also be qualitatively significant if the rms
dispersion with unit weighting (distributed according to a reduced
$\chi^2$ law) differs significantly from unity; that is, when
there is a basis to think that the model does not fit the
observational data well, or a strong correlation is suspected in
the observational data. The parameter uncertainties obtained using
the $\chi^2_P$ distribution can also be considered "external,"
since the dispersions of the "external" observational data are
used here.

The most conservative "external"\, parameter uncertainties are
obtained using the $\chi^2_M$ statistic, which does not assume the
model to be perfectly valid and uses the dispersions of the
observational data. It is also important that the same residual
(the $\chi^2_M$ statistic) is used both to search for the central
parameter values and to find their confidence intervals. However,
the confidence set $D$ can be empty for certain significance
levels $\alpha$ when using the $\chi^2_M$ statistic. As a result
of fitting the observations, only a less strong statement can be
made concerning the true parameter values, since we did not assume
by default that the model is perfectly valid when using the
$\chi^2_M$ statistic; namely, we can only indicate the probability
(the significance level $\alpha$) that wemistakenly reject the
hypothesis that the combined true parameter values belong to a
certain area $D$ (rather than, strictly speaking, the probability
$\gamma$ with which the confidence area $D$ encompasses this
combination of parameters). If we suppose that the model fits the
observational data, the probability for the area $D$ to encompass
the exact values of all the parameters is $\gamma = 1 - \alpha$,
where the probabilty $\gamma = 1$ is referred to as the confidence
level for the confidence area $D$. If, however, the model does not
fit the observational data, this may mean that we make a
second-order error in accepting the model (i.e., the model is
false, but is accepted according to the statistical criterion). In
this case, the equality $\gamma = 1 - \alpha$ is only approximate
for the confidence area.

However, we must remember that, in most cases, we are dealing with
a specific realization (a sample) of a light curve rather than the
entire observed light curve (a general population). Therefore,
difficulties that arise when trying to demonstrate the adequacy of
a model may be associated with considerable statistical deviations
of a specific observational data, rather than shortcomings of the
model itself. A light curve obtained at another epoch may match
the model adequately, with no difficulties arising when fitting
the observations. Therefore, strictly speaking, a specific
realization used in fitting does not necessarily imply that the
model is inadequate; it implies only that the available
observational data are not sufficient to ascertain whether the
model is adequate or not. It is this that enables us to estimate
the model parameters and their uncertainties even when the reduced
$\chi^2$, $\chi^2_{red}$, is significantly greater than unity.

It follows from these considerations that searches for model
parameters and their uncertainties in such problems should be
accompanied by testing of the model adequacy. The $\chi^2_M$
statistic can be used to test the adequacy of a model and verify
how it fits the observational data (since the model is not assumed
to be perfectly valid). Here, we must bear in mind that we accept
the model according to a statistical criterion not because it is
perfectly valid, but because there is no reason to reject it.

We can judge how well the model fits the obesrvational data
(whether the model is adequate enough) by estimating the
significance level $\alpha = \alpha_0$ starting from which the
model can be rejected (the value $\alpha_0$ is called the critical
value of the significance level). The higher the critical value
$\alpha_0$, the greater the probability that we make a first-order
error when rejecting a model (that is, the weaker the basis for us
to reject the model). The critical significance level $\alpha_0$
is associated with the reduced $\chi^2$ for the minimal residual,
$\chi^2_{red}=\frac{(\chi^2_{M-P})_{min}}{M-P}$ , which decreases
with increasing $\alpha_0$ \cite{Abubekerov2009}.

Strictly speaking, the residual minimized over nonlinear
parameters in nonlinear parametric problems is not distributed
according to a $\chi^2_{M-P}$ law, and only asymptotically
approaches this law as $M \rightarrow \infty$
\cite{Cherepashchuk1993}. Therefore, the criterion for an adequate
model that the $\chi^2_{red}$ red is close to unity can be used
only when the number of observational points $M$ is sufficiently
large. It was indicated in \cite{Abubekerov2009} that, if
$\chi^2_{red} = 1$, the critical significance level is $\alpha_0
\simeq 50\%$, which corresponds to a very good model. In fact, we
have the possibility in this case (when choosing a corresponding
significance level $\alpha$) to make up to 50\% first-order errors
in fitting if we reject the model; that is, we can err in every
second case in rejecting the model. Hence, we have no reason to
reject the model, and it can be accepted. When $\chi^2_{red}$
exceeds unity, the corresponding critical significance level is
$\alpha_0 < 50\%$. In this case, we do not make a large number of
firstorder errors in rejecting the model; i.e., we are correct in
most cases when we reject the model, and we therefore have a basis
to reject the model. If, however, we obtain $\chi^2_{red} < 1$,
the corresponding significance level is $\alpha_0 > 50\%$. When
the observational data are normally distributed, this situation is
very unlikely, even for a model that is perfectly valid (see, for
example, \cite{Abubekerov2009}); therefore, the value
$\chi^2_{red} < 1$, and the corresponding value $\alpha_0 > 50\%$,
can indicate the presence of a correlation in the input
observational data, which may include not only random but also
systematic errors.

Finally, let us illustrate again why it is necessary to verify the
model adequacy when deriving parameters from fits. An independent
way of verifying the model adequacy follows from the fact that the
number of observational points $M$ usually greatly exceeds the
number of parameters $P$, i.e., as a rule, the problem is highly
overdetermined. Strictly speaking, one should have an exact light
curve corresponding to the true parameter values to estimate their
uncertainties. However, an exact light curve (corresponding to the
perfect parameter values) is not available in reality; we have
only an observed light curve, subject to observational
uncertainties. The question thus arises of how well the observed
light curvematches the perfect light curve. In other words, before
finding the solution from a fit, we must answer the question of
whether we have a sufficient basis to substitute the observed
light curve for an exact light curve that is unknown a priori.

We can answer this fundamentally important question after checking
the model's adequacy. If $\chi^2_{red} \simeq 1$, that is,
$\alpha_0 \simeq 50\%$, there is no reason to reject the model,
and the model can be accepted. An imagined perfectly exact light
curve would optimally pass through the observational points of the
real light curve, and we would have a firm basis to substitute the
observed light curve for the perfectly exact light curve. The
obtained values of the model parameters are close to their true
values. Moreover, since the observed light curve optimally matches
the perfectly exact curve and the contribution of systematic
errors to the "exact" curve is negligible, we can obtain reliable
uncertainty estimates that encompass the true parameter values
with the given probability $\gamma$. We can estimate the parameter
uncertainties using various methods, depending on the strictness
with which we wish to judge the solution. These include the
differential-correction or Monte-Carlo methods, which enable us to
obtain "internal" uncertainties, the confidence-area method based
on the $\chi^2_M$ statistic, which yields "external"
uncertainties, or the confidence-area method based on the
$\chi^2_P$ statistic (the intermediate case).

However,if we obtain $\chi^2_{red} > 1$ and, correspondingly,
$\alpha_0 < 50\%$, we cannot make an unambiguous judgement about
whether or not to accept the model. The modelmay prove to be "bad"
since we can adopt for the model only a significance level
$\alpha$ for which the probability to make a first-order error is
small; i.e., we must only rarely be mistaken when rejecting the
model. In this case, the statistical criterion suggests that the
model should be rejected, and we construct a new and more perfect
model. It is often the case in practice that $\chi^2_{red}
> 1$ and the
corresponding critical significance level is $\alpha_0 < 10\%$. In
this case, if we accept the model, we must remember that the
obtained parameters and their confidence intervals are subject to
systematic errors due to either the fact that the perfectly exact
light curve does not optimally pass through the observed light
curve, or the presence of a strong correlation in the
observational data (note that, in this case, we cannot
unambiguously judge the systematic error to be due to a
correlation in the observational data, in contrast to the case
when $\alpha_0 > 50\%$). Of course, the obtained model parameters
and their uncertainties will be less reliable in these cases. The
questions raised above are considered in more detail elsewhere
\cite{Abubekerov2010, Abubekerov2008, Abubekerov2009,
Cherepashchuk1993}.

\section*{\normalsize OBSERVATIONAL DATA}

We analyze here a multicolor transit light curve of the binary
system HD 189733, which contains an exoplanet \cite{Pont2007}. The
light curves were obtained with the Hubble Space Telescope during
three observing runs, on May 22 and 26 and July 14, 2006, each
lasting five orbital turns of the space observatory.

The observations of HD 189733 were carried out using the Advanced
Camera for Surveys (ACS) in the HRC mode. In total, ten light
curves with eclipses were obtained in the ranges
$\lambda\lambda=5500-6000\AA$, $6000-6500\AA$, $6500-7000\AA$,
$7000-7500\AA$, $7500-8000\AA$, $8000-8500\AA$, $8500-9000\AA$,
$9000-9500\AA$, $9500-10000\AA$, $10000-10500\AA$. More detailed
information about the observational data is available in [5]. When
analyzing the light curves, we adopted the central wavelengths
$\lambda\lambda=5750\AA$, $6250\AA$, $6750\AA$, $7250\AA$,
$7750\AA$, $8250\AA$, $8750\AA$, $9250\AA$, $9750\AA$, $10250\AA$.

\begin{figure*}[h!]
\vspace{0cm} \epsfxsize=0.99\textwidth
\epsfbox{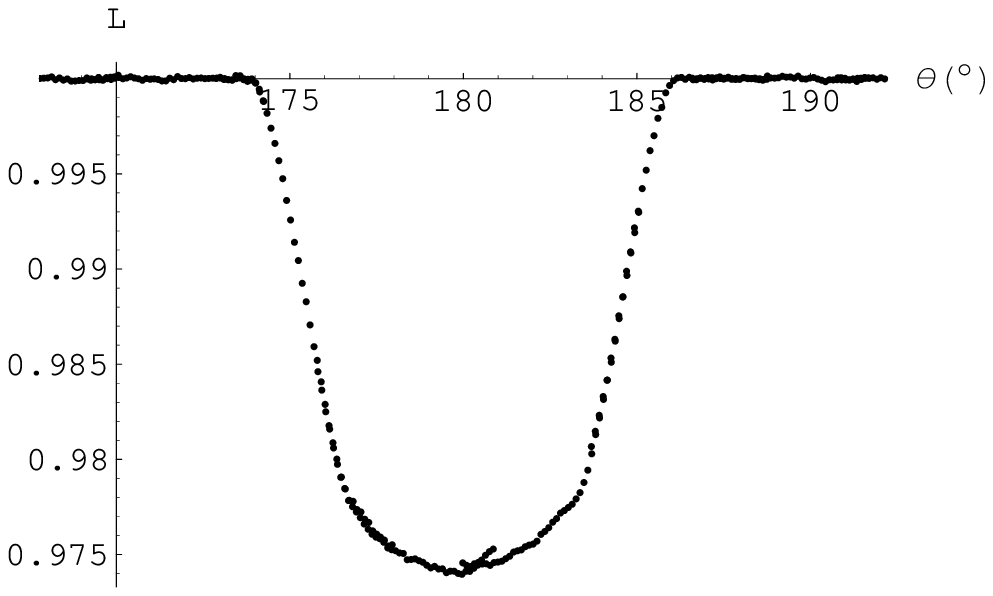} \caption{Light curve 1 of the
star HD 189733 in the filter $\lambda=5500-6000$ taken from
\cite{Pont2008}. The light curve has truncated wings and a
brightness jump at the minimum (due to a spot on the surface of
the star).} \label{LClam5500necorecct}
\end{figure*}

Each light curve contains 675 individual brightness estimates.
Figure \ref{LClam5500necorecct} shows the observed light curve in
the wavelength range $\lambda=5500-6000\AA$ (central wavelength
$\lambda = 5750\AA$).

We can see from Fig.\ref{LClam5500necorecct} that spots on the
surface of the star \cite{Pont2007} give rise to considerable
brightness changes at eclipse phases. There is an appreciable
shift in brightness between the right and left parts of the light
curve due to either a spot on the surface of the star or
systematic errors in the observations. These features required
careful analysis and some corrections to the light curve.

We assumed that the observational uncertainties could be described
by a normal distribution. The dispersions $\sigma^2$ of the
individual points in the light curve were assumed to be the same
for all points in the light-curve section used for the fitting.
The dispersions were determined by averaging the squared
differences between the observed points outside eclipse and
themean brightness outside eclipse, using the left and right
branches of the light curve both together and separately. We
ascribed the resulting values of $\sigma$ as the rms deviations of
the individual brightness measurements when fitting the
corresponding part of the light curve.

\renewcommand{\tablename}{Table}
\begin{table}[h!]
\caption{rms deviations and the brightness outside eclipse for the light curves}
\label{sigma_obs_tabl}\vspace{3mm} \centering 
\scriptsize
\begin{tabular}{c|c|c|c|c|c|c|c|c|c}
$\lambda (\AA)$ & $\L_{both}$ & $\sigma_{both}$ & $M_{both}$ & $L_{left}$ & $\sigma_{left}$& $M_{left}$ & $L_{right}$ & $\sigma_{right}$ & $M_{right}$ \\
\hline
$5750$   &  0.99998229   & 0.00007672 & 146 &  0.99997273 & 0.000078  & 79  & 0.99999736  & 0.000065  & 67 \\
$6250$   & 0.99997042   & 0.0003168    & 146 &  0.99994661 & 0.00035    & 79  & 1.0000079    & 0.00026    & 67 \\
$6750$   & 0.99994702   & 0.0002085    & 145 &  0.9999172   & 0.00022    & 78  & 0.99999414  & 0.00017    & 67 \\
$7250$   & 0.99996076   & 0.0001496    & 145 &  0.99993531 & 0.00016    & 78  & 1.000001      & 0.0001      & 67 \\
$7750$   & 0.99997562   & 0.0001421    & 146 &  0.99996303 & 0.00014    & 79  & 0.99999546  & 0.00014    & 67 \\
$8250$   & 0.99999351   & 0.0001321    & 147 &  0.99999245 & 0.00013    & 80  & 0.99999517  & 0.00013    & 67 \\
$8750$   & 1.0000042   & 0.0001558      & 147 &  1.0000095   & 0.00016    & 80  & 0.9999958    & 0.00015    & 67 \\
$9250$   & 0.99999268   & 0.0001766    & 147 &  0.99999       & 0.00018    & 80  & 0.9999969    & 0.00017    & 67 \\
$9750$   & 1.0000026   & 0.0002123      & 145 &  1.0000063   & 0.00022    & 78  & 0.99999678  & 0.0002      & 67 \\
$10250$ & 1.0000438   & 0.0003353      & 148 &  1.0000711   & 0.00036    & 80  & 1.0000006    & 0.00028    & 68 \\
\hline
\end{tabular}
\end{table}

We excluded part of the first observing run outside the eclipse at
phases $\theta<160^{\circ}$, where an obvious systematic shift was
observed relative to all other brightness measurements outside the
eclipse.

The mean brightness $L$ outside the eclipse, rms deviations of the
individual brightness measurements $\sigma$, and the number of
observational points in the eclipse $M$ for each light curve
(labeled "both" for the left and right branches together, "left"
for the left branch alone, and "right" for the right branch alone)
are given in Table \label{sigma_obs_tabl}. The accuracy of the
observed light curve is $\sim 10^{-4}$, which is $\sim 1\%$ of the
eclipse depth.

The main feature of the observed eclipse light curves is the
distortion at the minimum (at phases
$180^{\circ}<\theta<190^{\circ}$ for the points from the first
observing run), due to spots on the surface of the star
\cite{Pont2007}. Moreover, even if this effect is excluded from
consideration, the light curve remains appreciably distorted near
the minimum at $\theta \sim 180^{\circ}$. Therefore, in addition
to fitting using all the points in the light curve, we also
analyzed the left (descending) branch of the light curve without
$\theta>180^{\circ}$ and the right (ascending) branch without
points $\theta<180^{\circ}$.

\begin{figure*}[h!]
\vspace{0cm} \epsfxsize=0.99\textwidth
\epsfbox{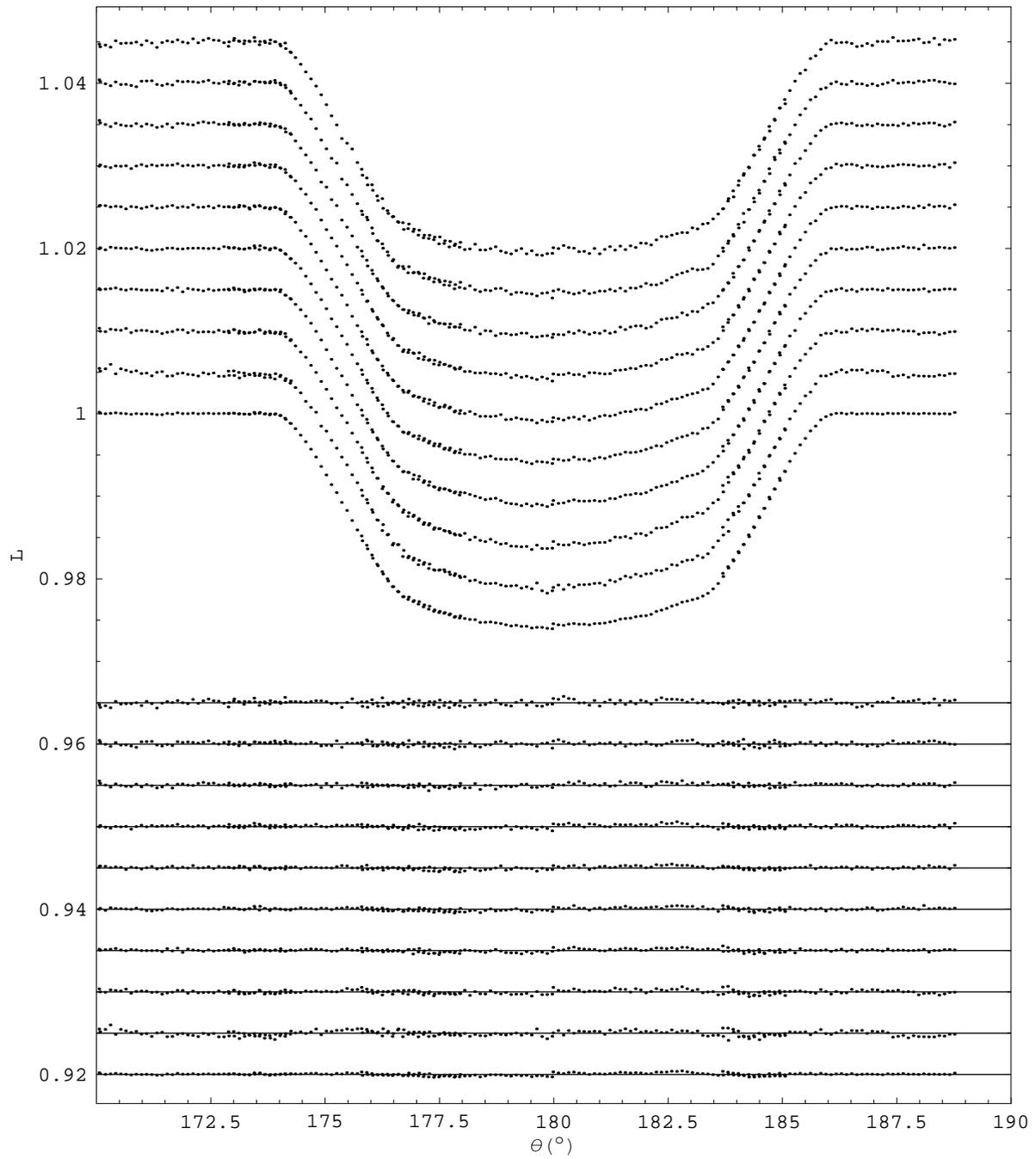} \caption{Light
curves of HD 189733 at $\lambda=5500-10500$. The wavelenght
increases upward. The residuals for the quadratic limb-darkening
law with the best-fit parameter values are given in the lower
part.} \label{LCnecorecct}
\end{figure*}

The resulting light curves that were analyzed are presented in
\ref{LCnecorecct}. For ease of viewing, the zero points of the
light curves for different $\lambda$ are shifted relative to each
other. The theoretical light curves and the corresponding residual
curves calculated using the best-fit model with quadratic
limb-darkening are shown in the same figure.

\section*{\normalsize FITTING OF THE LIGHT CURVES USING
THE LINEAR LIMB-DARKENING LAW}

The parameters derived from the observed light curves are the
radius of the star $r_s$, radius of the exoplanet $r_p$, orbital
inclination $i$, and coefficient $x$ in the linear limb-darkening
law. The HD 189733 system contains a star of spectral type K2V
\cite{Bouchy2005}. The orbital period was taken to be
$P_{orb}=2^{d}.218581$ \cite{Pont2007}, the planet-to-star mass
ratio to be $q=m_p/m_s=0.014$ \cite{Pont2007}, the orbit of the
system to be circular, and the radius of the relative orbit to be
unity. The residual was minimized simultaneously over all
parameters. Observational points from the phase interval covering
only the eclipsed part of the light curve were used (the numbers
of such points M for corresponding parts of the light curves are
given in Table \ref{sigma_obs_tabl}). The parts of the light
curves outside the eclipse were not used in the fitting, because
they were used to independently determine the rms deviations of
the points in the light curves (Table \ref{sigma_obs_tabl}).

\subsection*{\it\normalsize Differential-Correction Method}

The results of fitting of the left and right parts of the transit
light curve and the light curve as a whole are given in Tables
\ref{tabl_difflin_left}, \ref{tabl_difflin_right}, and
\ref{tabl_difflin}, respectively. The central parameter values and
their $2\sigma$ uncertainties were obtained using the
differential-correction method. Since we used a model that most
often proved to be "bad" (see below), we must provide $2\sigma$
parameter uncertainties, corresponding to $\gamma=95.5\%$.
Moreover, when we determine the parameter uncertainties using the
differential-correctionmethod in a multiparametric model, the true
value of each parameter falls into the corresponding error
interval with the given probability regardless of whether the true
values of the other parameters fall into their error intervals.

The question thus arises: what is the probability of all the
parameters falling into their error intervals? It is natural that
we want this probability to be at least at the specified
confidence level. If we consider the combined central parameter
values to be a P-dimensional random value varying in the
P-dimensional parameter space $\beta_1\ldots\beta_P$ , then, when
the true value $\bar{\beta_i}$ of one parameter $\beta_i$ falls
into the corresponding error interval, the point associated with
the combined true values will fall into the area between the two
(P-1)-dimensional hyperplanes $\beta_i = \beta_i^c-\kappa\sigma_i$
and $\beta_i = \beta_i^c+\kappa\sigma_i$ ($\kappa$ is a
coefficient corresponding to the chosen confidence level).

Therefore, the differential-correctionmethod is essentially
equivalent to finding a P-dimensional confidence area of this
type. When the true parameter values fall into their corresponding
error intervals, the point associated with the combined true
parameter values falls into the P-dimensional parallelepiped
centered on the point $\beta_1^c,\ldots,\beta_P^c$ and with sides
of length $2\kappa\sigma_1,\ldots,2\kappa\sigma_P$. It is obvious
that the probability for the combined true values to fall into
this parallelepiped will be less than the probability for the true
parameter values to independently fall into their error intervals,
that is, less than the given probability. To ensure that a point
in the P-dimensional parameter space corresponding to the combined
true parameter values falls into the P-dimensional parallelepiped
centered on the point $\beta_1^c,\ldots,\beta_P^c$ and with sides
proportional to $\sigma_1,\ldots,\sigma_P$ with the given
probability $\gamma$ (i.e., to achieve a situation where the true
parameter values all fall simultaneously into their corresponding
intervals with the given probability $\gamma$), we must choose the
intervals (dimensions of the P-dimensional parallelepiped) to be
$\mathrm{k} \kappa\sigma_i$, where $\mathrm{k}> 1$.

The probability for a P-dimensional random point to fall into the
P-dimensional parallelepiped is equal to the P-fold integral of
the multidimensional density distribution over the specified
parallelepiped. It is not only the rms deviations of the random
values $\beta_1^c,\ldots,\beta_P^c$ that we need to know before
defining the type of density function (and determining the
integral of the function). If a model is a linear function of the
parameters $\beta_1,\ldots,\beta_P$ (which have the
Gaussiandistributed central values $\beta_1^c,\ldots,\beta_P^c$)
\cite{Abubekerov2008}), this distribution function will be written
where $K$ is a normalizing coefficient and the matrix $A$ is the
inverse of the covariance matrix
$\mathbf{cov}(\beta_i^ñ,\beta_j^ñ)$ \footnote{The exponent in
(\ref{PDensity}) is the difference between the residual functional
as a function of the parameters and the minimum value of this
functional \cite{Abubekerov2008}, and each level surface of the
function $\mathfrak{f}$ (which is a P-dimensional ellipsoid)
bounds a confidence area obtainedwith a certain confidence level
using the $\chi_P^2$ distribution.}

\begin{equation}\label{PDensity}
\mathfrak{f}(\beta_1,\ldots,\beta_P) = K
e^{\displaystyle{\,\,\sum\limits_{i,j=1}^P A_{ij}(\beta_i -
\bar{\beta_i})(\beta_j - \bar{\beta_j})}}
\end{equation}

Therefore, even if we use a linear model (or a linear
approximation) to determine the probability for the true parameter
values to jointly fall into their error intervals (or if we wish
to solve for the coefficient $\mathrm{k}$), we must know the
covariances of the central values. The same values of the rms
deviations $\sigma_1,\ldots,\sigma_P$ can correspond to different
joint probabilities (different coefficients $\mathrm{k}$) in
different cases. In any case, however, $\mathrm{k} \kappa\sigma_i$
will be less than the corresponding projections of the confidence
area obtained using the $\chi_P^2$ distribution for the given
$\gamma$\footnote{Since these projections are dimensions of the
$P$-dimensional parallelepiped that are proportional to
$\sigma_1,\ldots,\sigma_P$, the probability for them to encompass
the true parameter values is greater than $\gamma$.}

For example, for  $\gamma = 0.955..$ (that is, with $\kappa = 2$),
the projection of the confidence area obtained using the
$\chi_P^2$ distribution is $\Delta_P = 3.12 \sigma$ for a
four-parameter model and $\Delta_P = 3.36 \sigma$ for a
five-parameter model \cite{Abubekerov2008}. Hence, at the
"$2\sigma$" ($\gamma = 0.955..$) confidence level, $\mathrm{k}
\lesssim 1.56$ for a four-parameter linear model and $\mathrm{k}
\lesssim 1.68$ for a fiveparameter linear model. Note that
$\mathrm{k} \lesssim 2.17$ for a fourparameter model and
$\mathrm{k} \lesssim 2.42$ for a five-parameter model at the 1ó
confidence level ($\gamma = 0.68$).

In addition to the parameter values,
table.\ref{tabl_difflin_left}, table.\ref{tabl_difflin_right},
table.\ref{tabl_difflin} also contain the minimum reduced $\chi^2$
values $\chi^2_{red}=(\chi^2_{M-P})_{min}/(M-P)$, and the critical
significance levels $\alpha_0$. As was already noted, the
$\chi^2_{red}$ values can be used to test whether the model
adequately fits the observational data. This indicates (Tables
\ref{tabl_difflin_left}, \ref{tabl_difflin_right},
\ref{tabl_difflin}) that our model was "bad" in most cases. When
fitting the right branch of the light curve (Table
\ref{tabl_difflin_right}).

Fitting of the left branch of the light curve (Table
\ref{tabl_difflin_left}) yielded $\chi^2_{red}<1$ for three of ten
wavelengths, with the corresponding $\alpha_0 > 50\%$. This
suggests that systematic errors strongly affect the left branch of
the light curve. As was noted above, it is these difficulties
(($\chi^2_{red} < 1$ and $\chi^2_{red}$) with fitting the light
curves of the HD 189733 system that force us to adopt the
confidence level $\gamma = 95.5\%$ (which corresponds to $2\sigma$
in the differential-correction method) rather than $68\%$,
indicative of "good" models. For the same reason, we were not able
to construct "exact" confidence areas D using the $\chi^2_M$
distribution and to provide the most conservative estimates of the
"external" parameter uncertainties for the right branch of the
light curve (Table \ref{tabl_difflin_right}) and for the entire
light curve (Table \ref{tabl_difflin}). We restricted our
consideration to producing asymptotic confidence areas for the
parameters using the $\chi^2_P$ statistic. We were able to
construct the confidence areas for the left branch of the light
curve using both the $\chi^2_P$ and the $\chi^2_M$ distributions.

\renewcommand{\tablename}{Table}
\begin{table}[h!]
\caption{Fitting of the left branches of the observed light curves
for HD 189733 \cite{Pont2008} with the linear limb-darkening law.
(Parameter uncertainties estimated using the
differential-correction method are given at the $2\sigma$ level.
The two last columns give the reduced $\chi^2$ and the
corresponding $\alpha_0$.)} \label{tabl_difflin_left} \vspace{3mm}
\centering \tiny
\begin{tabular}{c|c|c|c|c|c|c|c|c|c|c}
$\lambda (\AA)$ & $r_s^c$ & $2\sigma
_{\text{est}}\left(r_s^c\right)$ & $r_p^c$ & $2\sigma
_{\text{est}}\left(r_p^c\right)$ & $i^c ({}^{\circ})$ &
$2\sigma _{\text{est}}\left(i^c\right)({}^{\circ})$ & $x^c$ & $2\sigma _{\text{est}}\left(x^c\right)$ & $\chi^2_{red}$ & $\alpha_0$\\
\hline
5750 & 0.11184 & 0.00054 & 0.01760 & 0.00011 & 85.715 & 0.054 & 0.555 & 0.017 & 2.3457    & $1.9\cdot 10^{-9}$ \\
6250 & 0.1130 & 0.0011 & 0.01790 & 0.00025 & 85.60 & 0.11 & 0.609 & 0.037 & 0.51644          & 0.99996 \\
6750 & 0.11185 & 0.00090 & 0.01765 & 0.00020 & 85.697 & 0.091 & 0.615 & 0.028 & 0.78673& 0.956 \\
7250 & 0.11133 & 0.00073 & 0.01751 & 0.00016 & 85.754 & 0.073 & 0.590 & 0.022 & 0.96149& 0.695 \\
7750 & 0.11172 & 0.00066 & 0.01758 & 0.00014 & 85.716 & 0.065 & 0.556 & 0.020 & 1.1113  & 0.345 \\
8250 & 0.11219 & 0.00073 & 0.01768 & 0.00015 & 85.693 & 0.072 & 0.540 & 0.023 & 1.5651  & 0.0031 \\
8750 & 0.11225 & 0.00072 & 0.01765 & 0.00015 & 85.682 & 0.071 & 0.519 & 0.023 & 1.0306  & 0.53 \\
9250 & 0.11186 & 0.00080 & 0.01756 & 0.00016 & 85.731 & 0.079 & 0.490 & 0.025 & 1.0506  & 0.48 \\
9750 & 0.1110 & 0.0010 & 0.01740 & 0.00021 & 85.79 & 0.10 & 0.479 & 0.033 & 1.1975          & 0.19\\
10250 & 0.1117 & 0.0013 & 0.01755 & 0.00026 & 85.75 & 0.13 & 0.497 & 0.041 & 0.69904      & 0.99 \\
\hline
\end{tabular}
\end{table}

\renewcommand{\tablename}{Table}
\begin{table}[h!]
\caption{Fitting of the right branches of the observed light
curves for HD 189733 \cite{Pont2008} with the linear
limb-darkening law. (Parameter uncertainties estimated using the
differential-correction method are given at the $2\sigma$ level.
The two last columns give the reduced $\chi^2$ and the
corresponding $\alpha_0$.)} \label{tabl_difflin_right}
\vspace{3mm} \centering \tiny
\begin{tabular}{c|c|c|c|c|c|c|c|c|c|c}
$\lambda (\AA)$ & $r_s^c$ & $2\sigma
_{\text{est}}\left(r_s^c\right)$ & $r_p^c$ & $2\sigma
_{\text{est}}\left(r_p^c\right)$ & $i^c ({}^{\circ})$ &
$2\sigma _{\text{est}}\left(i^c\right) ({}^{\circ})$ & $x^c$ & $2\sigma _{\text{est}}\left(x^c\right)$ & $\chi^2_{red}$ & $\alpha_0$\\
\hline
5750 & 0.11159 & 0.00069 & 0.01738 & 0.00015 & 85.793 & 0.072 & 0.578 & 0.023 & 4.2535 & 0 \\
6250 & 0.1124  & 0.0015 & 0.01761 & 0.00034 & 85.71 & 0.16 & 0.590 & 0.053 & 1.3380    & 0.075155 \\
6750 & 0.1118  & 0.0012 & 0.01746 & 0.00028 & 85.77 & 0.13 & 0.631 & 0.041 & 1.9672    & 0.000029131 \\
7250 & 0.11164 & 0.00085 & 0.01741 & 0.00019 & 85.791 & 0.091 & 0.617 & 0.028 & 2.6644 & $1.2811 \cdot 10^{-10}$ \\
7750 & 0.11173 & 0.00086 & 0.01740 & 0.00019 & 85.784 & 0.091 & 0.593 & 0.029 & 1.4333 & 0.030473 \\
8250 & 0.11177 & 0.00078 & 0.01742 & 0.00017 & 85.786 & 0.082 & 0.570 & 0.026 & 1.3065 & 0.098526 \\
8750 & 0.11185 & 0.00088 & 0.01738 & 0.00019 & 85.771 & 0.092 & 0.534 & 0.030 & 1.4180 & 0.035518 \\
9250 & 0.11104 & 0.00092 & 0.01725 & 0.00019 & 85.845 & 0.097 & 0.521 & 0.032 & 1.1435 & 0.31474 \\
9750 & 0.1108 & 0.0012 & 0.01727 & 0.00025 & 85.84 & 0.12 & 0.497 & 0.042 & 1.4684     & 0.021212 \\
10250 & 0.1110 & 0.0013 & 0.01726 & 0.00027 & 85.83 & 0.13 & 0.506& 0.045 & 0.84083     & 0.89432\\
\hline
\end{tabular}
\end{table}

\subsection*{\it\normalsize Confidence-Area Method}

We used the $\chi^2_P$ and $\chi^2_M$ distributions for the
confidence-area method. We took $\gamma = 95.5\%$, which
corresponds to $2\sigma$ for the differential-correction method,
where $\sigma$ is the rms deviation. The $\chi^2_M$ statistic was
used only for the left branch of the light curve. The fitting
results for the left branch of the light curve obtained using the
confidence-area method based on the $\chi^2_M$ statistic are given
in Table \ref{tabl_XiM_lin_left}.

\begin{table}[h!]
\caption{Joint fitting of the right and left branches of the
observed light curves for HD 189733 \cite{Pont2008} with the
linear limbdarkening law. (Parameter uncertainties estimated using
the differential-correctionmethod are given at the $2\sigma$
level. The two last columns give the reduced $\chi^2$ and the
corresponding $\alpha_0$).} \label{tabl_difflin} \vspace{3mm}
\centering \tiny
\begin{tabular}{c|c|c|c|c|c|c|c|c|c|c}
$\lambda (\AA)$ & $r_s^c$ & $2\sigma
_{\text{est}}\left(r_s^c\right)$ & $r_p^c$ & $2\sigma
_{\text{est}}\left(r_p^c\right)$ & $i^c ({}^{\circ})$ &
$2\sigma _{\text{est}}\left(i^c\right) ({}^{\circ})$ & $x^c$ & $2\sigma _{\text{est}}\left(x^c\right)$ & $\chi^2_{red}$ & $\alpha_0$\\
\hline
 5750   & 0.11192 & 0.00059 & 0.01755 & 0.00012 & 85.726 & 0.060 & 0.556 & 0.019 & 5.5168  & 0 \\
 6250   & 0.1129   & 0.0011   & 0.01780 & 0.00024 & 85.64   & 0.11   & 0.600 & 0.036 & 1.0071 & 0.55 \\
 6750   & 0.11204 & 0.00086 & 0.01761 & 0.00019 & 85.706 & 0.088 & 0.614 & 0.028 & 1.5321 & 0.00010 \\
 7250   & 0.11171 & 0.00068 & 0.01753 & 0.00015 & 85.740 & 0.070 & 0.591 & 0.022 & 1.9306 & $6 \cdot 10^{-10}$ \\
 7750   & 0.11196 & 0.00068 & 0.01756 & 0.00014 & 85.717 & 0.069 & 0.561 & 0.022 & 2.0916 & $10^{-12}$ \\
 8250   & 0.11215 & 0.00066 & 0.01760 & 0.00014 & 85.715 & 0.067 & 0.545 & 0.022 & 2.2529 & $4 \cdot 10^{-15}$ \\
 8750   & 0.11230 & 0.00073 & 0.01758 & 0.00015 & 85.695 & 0.073 & 0.515 & 0.024 & 2.0264 & $2 \cdot 10^{-11}$ \\
 9250   & 0.11166 & 0.00070 & 0.01747 & 0.00014 & 85.758 & 0.071 & 0.494 & 0.023 & 1.4922 & 0.00029 \\
 9750   & 0.11103 & 0.00080 & 0.01736 & 0.00016 & 85.804 & 0.080 & 0.482 & 0.026 & 1.3541  & 0.0063 \\
 10250 & 0.11150 & 0.00096 & 0.01745 & 0.00020 & 85.765 & 0.097 & 0.493 & 0.032 & 0.79637 & 0.98 \\
\hline
\end{tabular}
\end{table}

The fitting results obtained using the confidencearea method based
on the $\chi_P^2$ distribution are given in Tables
\ref{tabl_XiP_lin_left}, \ref{tabl_XiP_lin_right},
\ref{tabl_XiP_lin}. Here, the best-fit values of $r_p$, $r_s$,
$i$, $x$ are given together with their uncertainties,
characterized by the projections of the asymptotic confidence area
$D$ in the four-parameter space onto the $r_p$, $r_s$, $i$, $x$
axes (the confidence intervals). To facilitate comparison with the
uncertainties obtained in the differential-correction method
(Tables \ref{tabl_difflin_left} - \ref{tabl_difflin}), Tables
\ref{tabl_XiP_lin_left} -- \ref{tabl_XiP_lin} show the values
$\Delta_P$, equal to half the confidence intervals. The tables for
the confidencearea method present parameter values corresponding
to the middles of the confidence intervals, rather than with the
central parameter values. The probability that the exact value of
each parameter $r_p$, $r_s$, $i$, $x$ is encompassed within the
corresponding confidence interval exceeds 95.5\%. The probability
that the exact solution (the combination of all parameters $r_p$,
$r_s$, $i$, $x$) is encompassed within the asymptotic confidence
area is close to the given probability $\gamma = 95.5\%$, since
the number of points in the light curve is large ($M \gtrsim 70$).
The probability that the exact solution is simultaneously
encompassed within all confidence intervals exceeds the given
$\gamma = 95.5\%$.

\begin{table}[h!]
\caption{Fitting of the left branches of the observed light curves
for HD 189733 \cite{Pont2008} with the linear limb-darkening law.
(Parameter uncertainties were estimated using the confidence-area
method based on the $\chi^2_P$ distribution, with
$\gamma=0.955$.)} \label{tabl_XiP_lin_left} \vspace{3mm}
\centering \small
\begin{tabular}{c|c|c|c|c|c|c|c|c}
$\lambda (\AA)$ & $r_s$ & $\Delta _P\left(r_s\right)$ & $r_p$ &
$\Delta _P\left(r_p\right)$ & $i$ & $\Delta
_P\left(i\right)$ & $x$ & $\Delta_P\left(x\right)$ \\
\hline
5750 & 0.11184 & 0.00055 & 0.01760 & 0.00011 & $85.715^{\circ}$  & $0.054^{\circ}$ & 0.555 & 0.017 \\
6250 & 0.1130 & 0.0025 & 0.01788 & 0.00055 &   $85.61^{\circ} $ &  $0.25^{\circ}$ & 0.607 & 0.080 \\
6750 & 0.1118 & 0.0016 & 0.01764 & 0.00035 &   $85.70^{\circ} $ &  $0.16^{\circ}$ & 0.614 & 0.049 \\
7250 & 0.1113 & 0.0012 & 0.01751 & 0.00025 &   $85.76^{\circ} $ &  $0.12^{\circ}$ & 0.590 & 0.035 \\
7750 & 0.11171 & 0.00094 & 0.01758 & 0.00020 & $85.717^{\circ}$ &  $0.092^{\circ}$ & 0.556 & 0.029 \\
8250 & 0.11217 & 0.00091 & 0.01767 & 0.00019 & $85.695^{\circ}$ &  $0.090^{\circ}$ & 0.540 & 0.029 \\
8750 & 0.1123 & 0.0011 & 0.01765 & 0.00022 &   $85.68^{\circ} $ &  $0.10^{\circ}$ & 0.518 & 0.034 \\
9250 & 0.1119 & 0.0012 & 0.01756 & 0.00025 &   $85.73^{\circ} $ &  $0.12^{\circ}$ & 0.490 & 0.040 \\
9750 & 0.1110 & 0.0015 & 0.01739 & 0.00029 &   $85.79^{\circ} $ &  $0.14^{\circ}$ & 0.477 & 0.046 \\
10250 & 0.1117 & 0.0026 & 0.01754 & 0.00053 &  $85.75^{\circ} $ &  $0.26^{\circ}$ & 0.493 & 0.082 \\
\hline
\end{tabular}
\end{table}

\begin{table}[h!]
\caption{Fitting of the right branches of the observed light
curves for HD 189733 \cite{Pont2008} with the linear
limb-darkening law. (Parameter uncertainties were estimated using
the confidence-area method based on the $\chi^2_P$ distribution,
with $\gamma=0.955$.)} \label{tabl_XiP_lin_right} \vspace{3mm}
\centering \small
\begin{tabular}{c|c|c|c|c|c|c|c|c}
$\lambda (\AA)$ & $r_s$ & $\Delta _P\left(r_s\right)$ & $r_p$ &
$\Delta _P\left(r_p\right)$ & $i$ & $\Delta
_P\left(i\right)$ & $x$ & $\Delta_P\left(x\right)$ \\
\hline
5750 & 0.11159 & 0.00052 & 0.01738 & 0.00011 & $85.794^{\circ}$ & $0.055^{\circ}$ & 0.578 & 0.018 \\
6250 & 0.1124 & 0.0019 & 0.01760 & 0.00043 &   $85.72^{\circ} $ & $0.20^{\circ} $ & 0.589 & 0.069 \\
6750 & 0.1118 & 0.0013 & 0.01745 & 0.00031 &   $85.77^{\circ} $ & $0.14^{\circ} $ & 0.631 & 0.046 \\
7250 & 0.11164 & 0.00082 & 0.01741 & 0.00019 & $85.792^{\circ}$ & $0.089^{\circ}$ & 0.617 & 0.028 \\
7750 & 0.1117 & 0.0011 & 0.01739 & 0.00025 &   $85.79^{\circ} $ & $0.12^{\circ} $ & 0.593 & 0.038 \\
8250 & 0.1118 & 0.0010 & 0.01742 & 0.00023 &   $85.79^{\circ} $ & $0.11^{\circ} $ & 0.570 & 0.036 \\
8750 & 0.1118 & 0.0011 & 0.01737 & 0.00024 &   $85.77^{\circ} $ & $0.12^{\circ} $ & 0.533 & 0.039 \\
9250 & 0.1110 & 0.0013 & 0.01725 & 0.00028 &   $85.85^{\circ} $ & $0.14^{\circ} $ & 0.520 & 0.046 \\
9750 & 0.1108 & 0.0015 & 0.01726 & 0.00031 &   $85.85^{\circ} $ & $0.16^{\circ} $ & 0.496 & 0.053 \\
10250 & 0.1110 & 0.0022 & 0.01725 & 0.00045 &  $85.83^{\circ} $ & $0.23^{\circ} $ & 0.503 & 0.076 \\
\hline
\end{tabular}
\end{table}

\begin{table}[h!]
\caption{Joint fitting of the right and left branches of the
observed light curves for HD 189733 \cite{Pont2008} with the
linear limbdarkening law. (Parameter uncertainties were estimated
using the confidence-area method based on the $\chi^2_P$
distribution, with $\gamma=0.955$).} \label{tabl_XiP_lin}
\vspace{3mm} \centering \small
\begin{tabular}{c|c|c|c|c|c|c|c|c}
$\lambda (\AA)$ & $r_s$ & $\Delta _P\left(r_s\right)$ & $r_p$ &
$\Delta _P\left(r_p\right)$ & $i$ & $\Delta
_P\left(i\right)$ & $x$ & $\Delta_P\left(x\right)$ \\
\hline
5750 & 0.11192 & 0.00039 & 0.017552 & 0.000083 & $85.726^{\circ}$ & $0.040^{\circ}$ & 0.556 & 0.013 \\
6250 & 0.1129 & 0.0017 & 0.01780 & 0.00036 &     $85.64^{\circ} $ & $0.17^{\circ} $ & 0.598 & 0.055 \\
6750 & 0.1120 & 0.0011 & 0.01761 & 0.00024 &     $85.71^{\circ} $ & $0.11^{\circ} $ & 0.614 & 0.035 \\
7250 & 0.11170 & 0.00078 & 0.01752 & 0.00017 &   $85.743^{\circ}$ & $0.080^{\circ}$ & 0.591 & 0.025 \\
7750 & 0.11194 & 0.00072 & 0.01755 & 0.00015 &   $85.719^{\circ}$ & $0.072^{\circ}$ & 0.562 & 0.023 \\
8250 & 0.11215 & 0.00068 & 0.01760 & 0.00014 &   $85.715^{\circ}$ & $0.069^{\circ}$ & 0.545 & 0.022 \\
8750 & 0.11229 & 0.00078 & 0.01758 & 0.00016 &   $85.698^{\circ}$ & $0.078^{\circ}$ & 0.515 & 0.026 \\
9250 & 0.11166 & 0.00093 & 0.01747 & 0.00019 &   $85.759^{\circ}$ & $0.095^{\circ}$ & 0.493 & 0.031 \\
9750 & 0.1110 & 0.0010 & 0.01736 & 0.00021 &     $85.81^{\circ} $ & $0.10^{\circ} $ & 0.482 & 0.034 \\
10250 & 0.1115 & 0.0017 & 0.01745 & 0.00036 &    $85.77^{\circ} $ & $0.18^{\circ} $ & 0.491 & 0.058 \\
\hline
\end{tabular}
\end{table}

\begin{table}[h!]
\caption{Fitting of the left branches of the observed light curves
for HD 189733\cite{Pont2008} with the linear limb-darkening law.
(Parameter uncertainties were estimated using the confidence-area
method based on the $\chi^2_M$ distribution, with
$\gamma=0.955$).} \label{tabl_XiM_lin_left} \vspace{3mm}
\centering \small
\begin{tabular}{c|c|c|c|c|c|c|c|c}
$\lambda (\AA)$ & $r_s$ & $\Delta _M\left(r_s\right)$ & $r_p$ &
$\Delta _M\left(r_p\right)$ & $i$ & $\Delta
_M\left(i\right)$ & $x$ & $\Delta_M\left(x\right)$ \\
\hline
6250 & 0.1127 & 0.0066 & 0.0178 & 0.0014   & $85.67^{\circ}$ & $0.67^{\circ}$ & 0.59 & 0.22 \\
6750 & 0.1118 & 0.0034 & 0.01761 & 0.00074 & $85.71^{\circ}$ & $0.34^{\circ}$ & 0.61 & 0.10 \\
7250 & 0.1113 & 0.0020 & 0.01750 & 0.00043 & $85.76^{\circ}$ & $0.20^{\circ}$ & 0.589 & 0.061 \\
7750 & 0.1117 & 0.0013 & 0.01758 & 0.00027 & $85.72^{\circ}$ & $0.13^{\circ}$ & 0.556 & 0.040 \\
8750 & 0.1122 & 0.0017 & 0.01765 & 0.00035 & $85.69^{\circ}$ & $0.17^{\circ}$ & 0.517 & 0.055 \\
9250 & 0.1118 & 0.0019 & 0.01755 & 0.00039 & $85.74^{\circ}$ & $0.19^{\circ}$ & 0.488 & 0.061 \\
9750 & 0.1110 & 0.0016 & 0.01739 & 0.00032 & $85.79^{\circ}$ & $0.16^{\circ}$ & 0.477 & 0.051 \\
10250 & 0.1116 & 0.0059 & 0.0175 & 0.0012 &  $85.78^{\circ}$ & $0.61^{\circ}$ & 0.47 & 0.19 \\
\hline
\end{tabular}
\end{table}

\section*{\normalsize THE PLANET'S RADIUS AND LINEAR
LIMB-DARKENING COEFFICIENT AS FUNCTIONS OF WAVELENGTH IN THE
LINEAR LAW}

Our model for the system HD 189733 with a linear limb-darkening
law proved to be "bad" for most wavelengths. This is not
surprising, since the light curves trace spots on the surface of
the star \cite{Pont2008}, which are not taken into account in the
model. Only for the left branch of the light curve can ourmodel
with linear limb-darkening be rejected for most wavelengths at a
fairly high significance level, so that we can estimate the most
conservative "external" parameter uncertainties for $\gamma =
95.5\%$ using the $\chi^2_M$ distribution (Table
\ref{tabl_XiM_lin_left}).

The exoplanet-to-star radius ratio $r_p/r_s$ in the HD 189733
system was obtained in \cite{Pont2008} as a function of
wavelength. Our results based on our analysis of the complete
light curves are given in Table \ref{Pont_sravnenie}, together
with the results obtained in \cite{Pont2008}. Since we are
searching for the individual parameters $r_p$ and $r_s$, rather
than their ratio $r_p/r_s$, the uncertainty in $r_p/r_s$ was
obtained using the differential-correctionmethod as an uncertainty
for a new parameter. The general formula for estimating the rms
deviations when changing variables is given in
\cite{Abubekerov2010}. In our case,

 $$
 \sigma(r^c_p/r^c_s) = \sqrt{\left(\frac{r^c_p}{{(r^c_s)}^2}\right)^2 \sigma^2(r^c_s) - 2 \mathbf{cov}(r^c_p,r^c_s) \left(\frac{r^c_p}{{(r^c_s)}^2}\right)
 \left(\frac{1}{r^c_s}\right) + \left(\frac{1}{r^c_s}\right)^2 \sigma^2(r^c_p)}
 $$

where $\mathbf{cov}(\cdot,\cdot)$ is the procedure for searching
for the covariances of random values. Our uncertainties are larger
than those of \cite{Pont2008}, where fits were obtained for the
geometrical parameters $r_s, r_p, i$, with the limb-darkening
coefficients being fixed. Since we fit the light curves allowing
both the geometrical parameters and the limb-darkening
coefficients to vary, the larger number of degrees of freedom in
our model yields larger parameter uncertainties.

\begin{figure*}[h!]
\vspace{0cm} \epsfxsize=0.99\textwidth \epsfbox{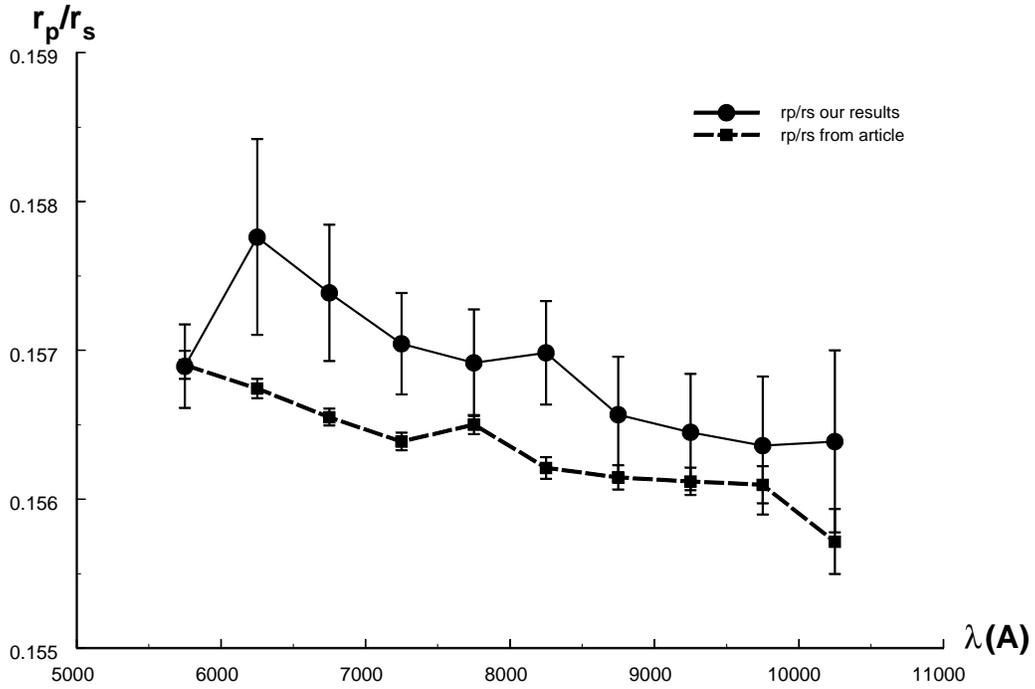}
\caption{The planet-to-star radius ratio as a function of
wavelength obtained in this paper (filled circles) and according
to \cite{Pont2008} (filled squares). The indicated uncertainties
are $1\sigma$. Since we do not specify the limb-darkening
coefficient, and are searching for it together with the other
parameters of the problem, our uncertainties are larger.}
\label{ris3}
\end{figure*}

Figure \ref{ris3} indicates a certain increase in the planet`s
radius with decreasing wavelength. This effect is significant at
the $1\sigma$ level, but not the $2\sigma$ level. Our values of
the exoplanet radius are systematically larger than those
determined in \cite{Pont2007} (by 0.3\%); this is due to the
different normalization procedures applied in the analysis of the
light curve and the fact that we searched for the limb-darkening
coefficient together with other parameters, rather than fixing
this coefficient.

\begin{table}[h!]
\caption{Comparison of the $r_p/r_s$ values as a function of
wavelength obtained in this paper and in \cite{Pont2008}. (The
$1\sigma$ uncertainties are given).} \label{Pont_sravnenie}
\vspace{3mm} \centering \small
\begin{tabular}{c|c|c}
$\lambda (\AA)$ & Íàøè ðàñ÷åòû & Èç ðàáîòû \cite{Pont2008} \\
\hline
$5750$   &  $0.156894 \pm0.000281$ & $0.156903 \pm0.000095$\\
$6250$   &  $0.157762 \pm0.000658$ & $0.156744 \pm0.000065$\\
$6750$    & $0.157387 \pm0.000458$ & $0.156552 \pm0.000057$\\
$7250$   &  $0.157045 \pm0.000341$ & $0.156388 \pm0.000059$\\
$7750$   &  $0.156917 \pm0.000358$ & $0.156501 \pm0.000064$\\
$8250$   &  $0.156984 \pm0.000348$ & $0.156210 \pm0.000073$\\
$8750$   &  $0.156569 \pm0.000388$ & $0.156147 \pm0.000081$\\
$9250$   &  $0.156451 \pm0.000391$ & $0.156120 \pm0.000092$\\
$9750$   &  $0.156361 \pm0.000463$ & $0.156097 \pm0.000125$\\
$10250$  & $0.156389  \pm0.000611$ & $0.155716 \pm0.000218$\\
\hline
\end{tabular}
\end{table}

\begin{table}[h!]
\caption{Fitting of the left branches of the observed light curves
for HD 189733 \cite{Pont2008} with the quadratic limb-darkening
law. (Parameter uncertainties estimated using the
differential-correction method are given at the $2\sigma$ level.
The two last columns present the values of the reduced $\chi^2$
and the corresponding critical significance levels $\alpha_0$).}
\label{tabl_diff_Knutson_nonlin_left} \vspace{3mm} \centering
\tiny
\begin{tabular}{c|c|c|c|c|c|c|c|c|c|c|c|c}
$\lambda (\AA)$ & $r_s^c$ & $2\sigma
_{\text{est}}\left(r_s^c\right)$ & $r_p^c$ & $2\sigma
_{\text{est}}\left(r_p^c\right)$ & $i^c ({ }^{\circ})$ &
$2\sigma _{\text{est}}\left(i^c\right) ({ }^{\circ})$ & $x_1^c$ & $2\sigma _{\text{est}}\left(x_1^c\right)$ & $y_1^c$ & $2\sigma _{\text{est}}\left(y_1^c\right)$ & $\chi^2_{red}$ & $\alpha_0$\\
\hline
5750 & 0.11207 & 0.00083 & 0.01754 & 0.00022 & 85.721 & 0.077 & 0.44 & 0.16 & 0.18 & 0.26 & 2.3115 & $8\cdot 10^{-9}$\\
 6250 & 0.1132 & 0.0020 & 0.01782 & 0.00050 & 85.62 & 0.17 & 0.50 & 0.37 & 0.16 & 0.60 & 0.52219 & 0.99997 \\
 6750 & 0.1120 & 0.0014 & 0.01759 & 0.00035 & 85.70 & 0.12 & 0.52 & 0.26 & 0.14 & 0.42 & 0.79144 & 0.96 \\
 7250 & 0.11144 & 0.00094 & 0.01748 & 0.00024 & 85.758 & 0.088 & 0.53 & 0.20 & 0.08 & 0.32 & 0.97038 & 0.70\\
 7750 & 0.11189 & 0.00091 & 0.01754 & 0.00023 & 85.720 & 0.083 & 0.47 & 0.20 & 0.13 & 0.30 & 1.1150  & 0.37 \\
 8250 & 0.1125 & 0.0022 & 0.01754 & 0.00048 & 85.72 & 0.16 & 0.32 & 0.24 & 0.33 & 0.48 & 1.5210   & 0.0074\\
 8750 & 0.1123 & 0.0010 & 0.01757 & 0.00026 & 85.704 & 0.094 & 0.42 & 0.23 & 0.14 & 0.35 & 1.0401 & 0.54\\
 9250 & 0.1121 & 0.0013 & 0.01748 & 0.00036 & 85.74 & 0.12 & 0.34 & 0.26 & 0.23 & 0.41 & 1.0466 & 0.53 \\
 9750 & 0.1112 & 0.0013 & 0.01737 & 0.00034 & 85.79 & 0.12 & 0.41 & 0.33 & 0.10 & 0.49 & 1.2103 & 0.198\\
 10250 & 0.1131 & 0.0044 & 0.0175 & 0.0012 & 85.71 & 0.29 & 0.05 & 0.43 & 0.65 & 0.45 & 0.66148 & 0.997 \\
\hline
\end{tabular}
\end{table}

Note that the difference between the exoplanet radius in HD 209458
obtained in the blue and the red does not seem to be significant
\cite{Abubekerov2010}: $\overline{r_p} = 0.0139\pm 0.0003$ for
$\overline{\lambda} = 3750$ (the mean value for $\lambda = 3201,
3750, 4300\AA\AA$) and $\overline{r_p}=0.0138\pm 0.0002$ for
$\lambda = 8732$ (the mean value for $\lambda = 7755, 8732,
9708\AA\AA$); the $2\sigma$ uncertainties are indicated.

\begin{table}[h!]
\caption{Fitting of the right branches of the observed light
curves for HD 189733 \cite{Pont2008} with the quadratic
limb-darkening law. (Parameter uncertainties estimated using the
differential-correction method are given at the $2\sigma$ level.
The two last columns present the values of the reduced $\chi^2$
and the corresponding critical significance levels $\alpha_0$).}
\label{tabl_diff_Knutson_nonlin_right} \vspace{3mm} \centering
\tiny
\begin{tabular}{c|c|c|c|c|c|c|c|c|c|c|c|c}
$\lambda (\AA)$ & $r_s^c$ & $2\sigma
_{\text{est}}\left(r_s^c\right)$ & $r_p^c$ & $2\sigma
_{\text{est}}\left(r_p^c\right)$ & $i^c ({ }^{\circ})$ &
$2\sigma _{\text{est}}\left(i^c\right) ({ }^{\circ})$ & $x_1^c$ & $2\sigma _{\text{est}}\left(x_1^c\right)$ & $y_1^2$ & $2\sigma _{\text{est}}\left(y_1^c\right)$ & $\chi^2_{red}$ & $\alpha_0$\\
\hline
5750 & 0.11094 & 0.00053 & 0.017531 & 0.000095 & 85.781 & 0.044 & 0.86 & 0.20 & -0.42 & 0.27 & 3.7576 & 0 \\
6250 & 0.1113 & 0.0011 & 0.01794 & 0.00016 & 85.681 & 0.078 & 1.13 & 0.44 & -0.78 & 0.58 & 1.1957 & 0.26 \\
6750 & 0.11080 & 0.00089 & 0.01765 & 0.00014 & 85.761 & 0.070 & 1.01 & 0.33 & -0.56 & 0.45 & 1.8492 & $2.6\cdot10^{-4}$ \\
7250 & 0.11115 & 0.00072 & 0.01754 & 0.00014 & 85.772 & 0.060 & 0.84 & 0.25 & -0.33 & 0.35 & 2.5861 & $1.27\cdot10^{-9}$ \\
7750 & 0.11109 & 0.00069 & 0.01752 & 0.00013 & 85.776 & 0.058 & 0.85 & 0.25 & -0.38 & 0.35 & 1.3392 & 0.089 \\
8250 & 0.11076 & 0.00050 & 0.017623 & 0.000080 & 85.778 & 0.040 & 0.99 & 0.20 & -0.61 & 0.27 & 0.99821 & 0.65 \\
8750 & 0.11116 & 0.00069 & 0.01753 & 0.00012 & 85.761 & 0.056 & 0.85 & 0.27 & -0.46 & 0.37 & 1.3138 & 0.11 \\
9250 & 0.11083 & 0.00089 & 0.01731 & 0.00018 & 85.839 & 0.079 & 0.63 & 0.28 & -0.16 & 0.40 & 1.1495 & 0.338 \\
9750 & 0.1120 & 0.0024 & 0.01735 & 0.00061 & 85.78 & 0.21 & 0.31 & 0.44 & 0.26 & 0.73 & 1.5716 & 0.0089\\
10250 & 0.1124 & 0.0033 & 0.01736 & 0.00076 & 85.74 & 0.27 & 0.26 & 0.48 & 0.33 & 0.87 & 0.88100 & 0.86 \\
\hline
\end{tabular}
\end{table}

\begin{figure*}[h!]
\vspace{0cm} \epsfxsize=0.99\textwidth
\epsfbox{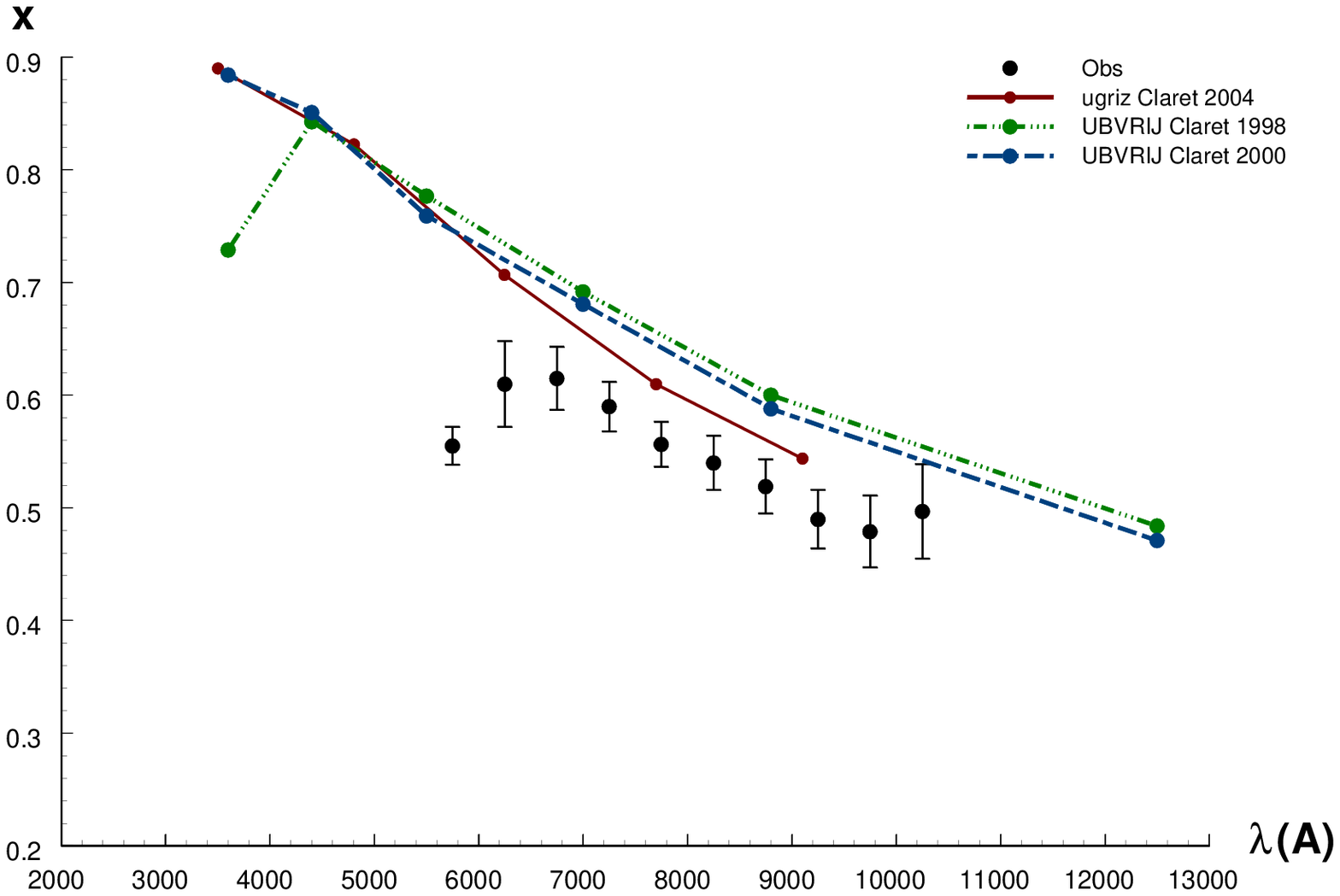}
\caption{Limb-darkening coefficient $x$ for HD 189733 as a
function of wavelength $\lambda$ derived for the linear
limb-darkening law and the left branches of the light curves from
\cite{Pont2008}. The uncertainties of the limb-darkening
coefficient were obtained using the differential-correction method
and are given at the 2$\sigma$ level. The theoretical values of
the limb-darkening coefficients in the ugriz and UBVRIJ
photometric systems were taken from
{Claret2004,Claret1998,Claret2000}.}
\label{LD_x1_diff_HD189733_left}
\end{figure*}

\begin{figure*}[h!]
\vspace{0cm} \epsfxsize=0.99\textwidth
\epsfbox{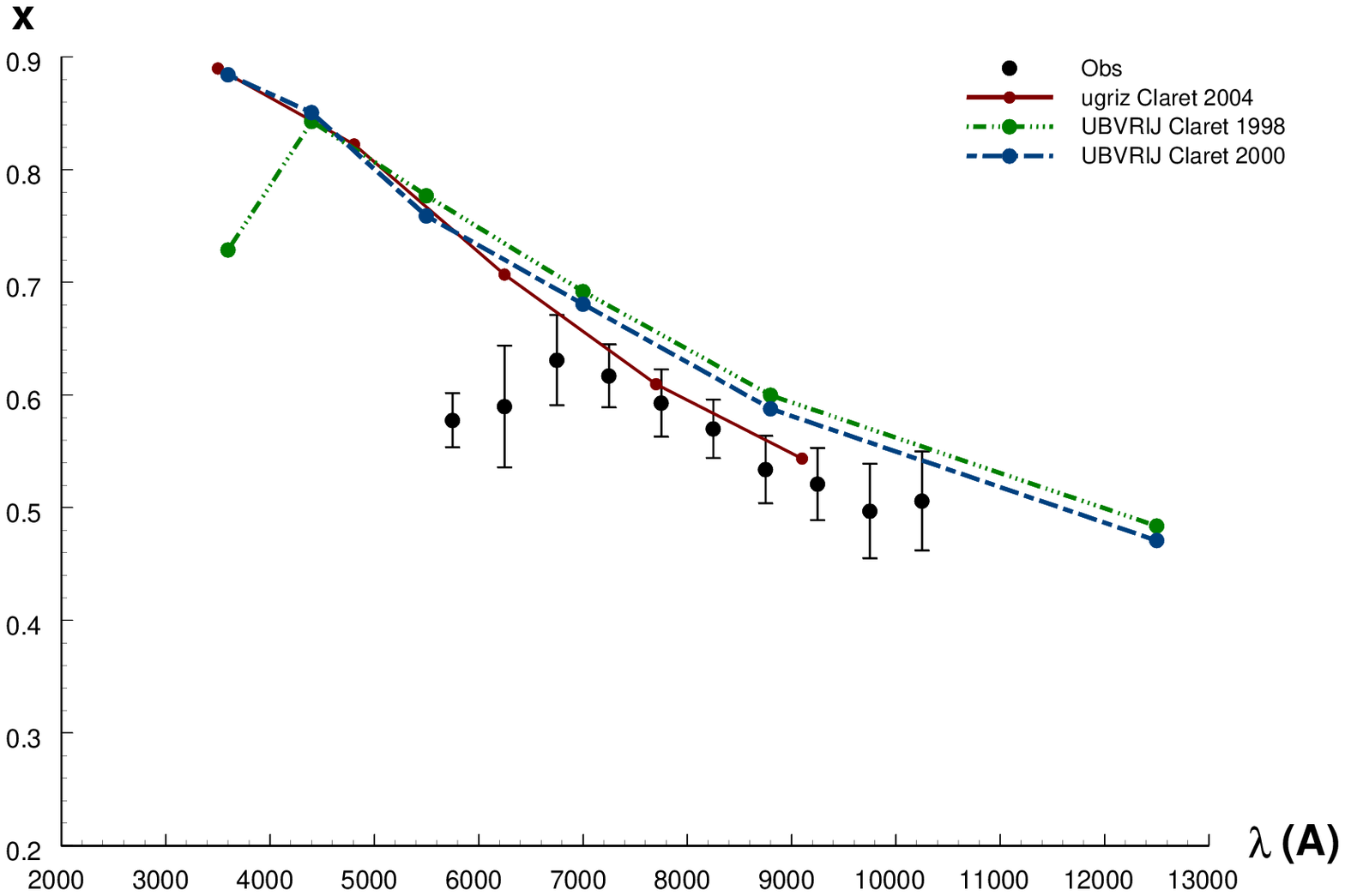}
\caption{Limb-darkening coefficient $x$ for HD 189733 as a
function of wavelength $\lambda$ derived for the linear
limb-darkening law and the right branches of the light curves from
[6]. The uncertainties of limb-darkening coefficient were obtained
using the differential-correction method and are given at the
2$\sigma$ level. The theoretical values of the limb-darkening
coefficient in the ugriz and UBV RIJ photometric systems were
taken from \cite{Claret2004,Claret1998,Claret2000}.}
\label{LD_x1_teor_obs_diff_HD189733_right}
\end{figure*}

\begin{figure*}[h!]
\vspace{0cm} \epsfxsize=0.99\textwidth
\epsfbox{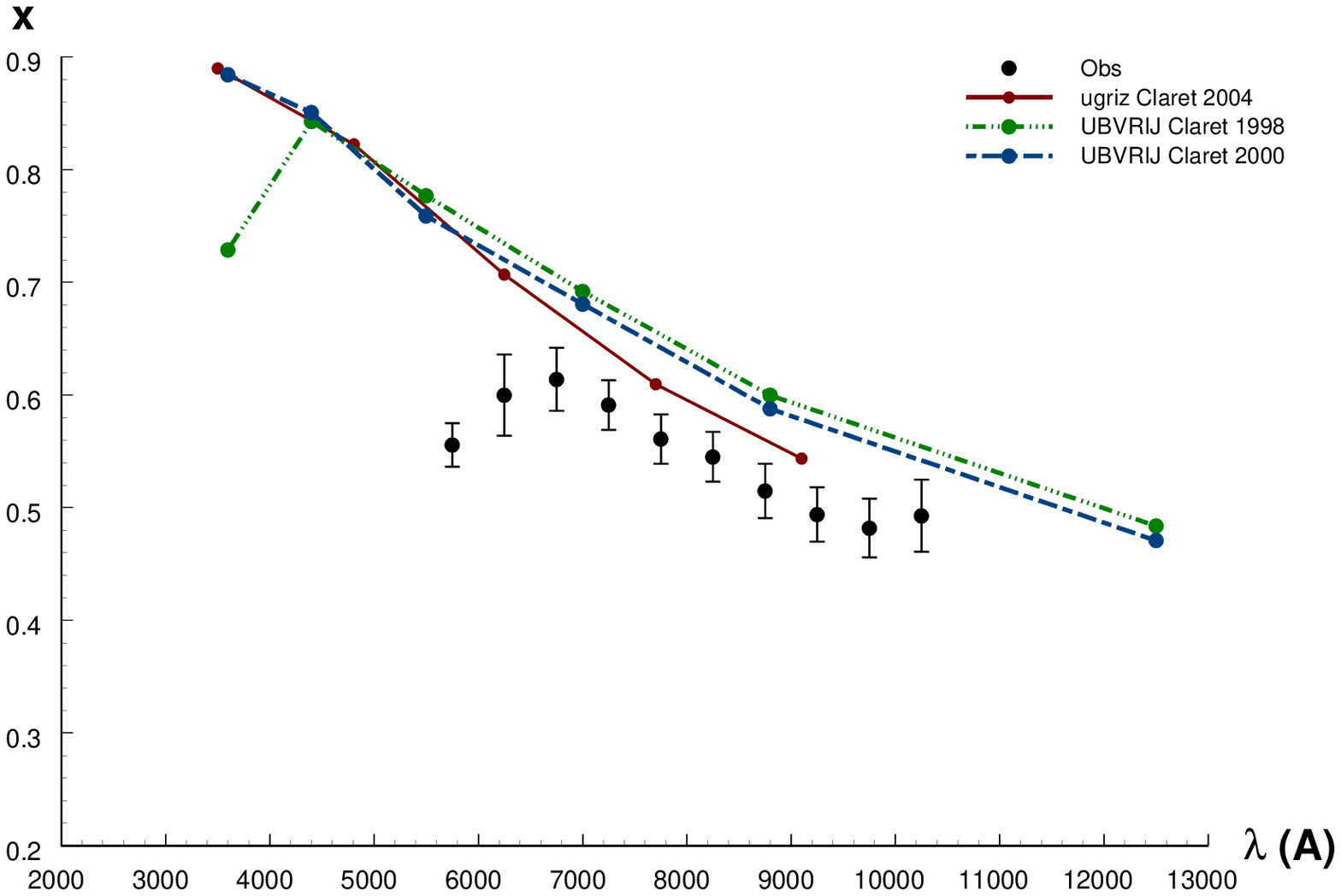} \caption{
Limb-darkening coefficient $x$ forHD189733 as a function of
wavelength $\lambda$ derived for the linear limb-darkening law and
the joint analysis of the right and left branches of the light
curves from \cite{Pont2008}. The uncertainties of limb-darkening
coefficient were obtained using the differential-correction method
and are given at the 2ó level. The theoretical values of the
limb-darkening coefficient in the ugriz and UBVRIJ photometric
systems were taken from  \cite{Claret2004,Claret1998,Claret2000}.}
\label{LD_x1_diff_HD189733_lr}
\end{figure*}

\begin{figure*}[h!]
\vspace{0cm} \epsfxsize=0.99\textwidth
\epsfbox{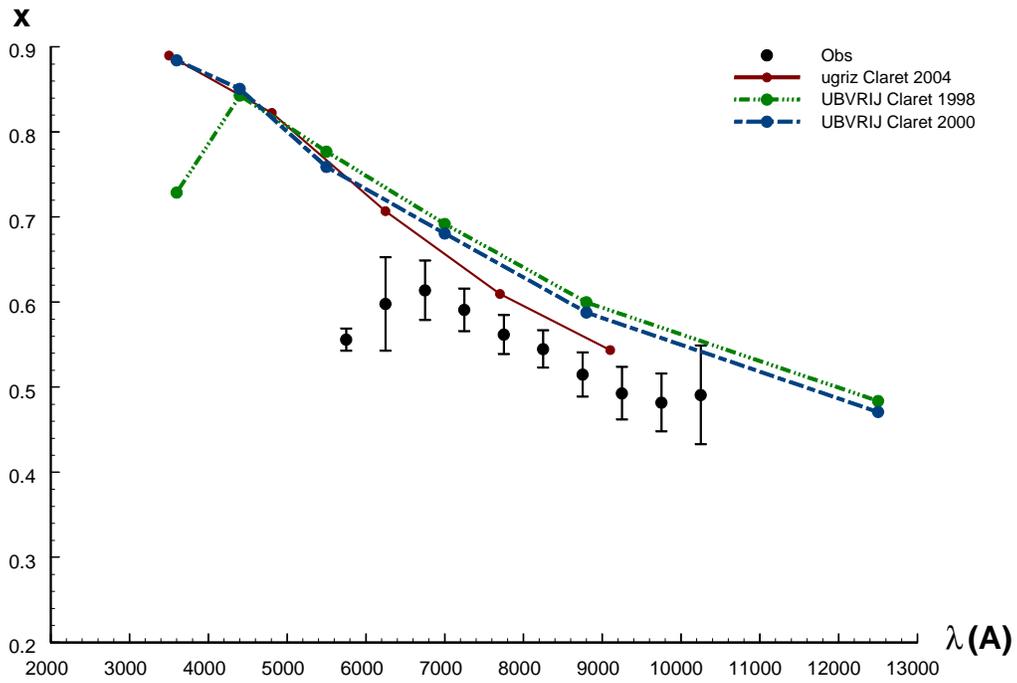} \caption{Same
function as in Fig.6 but with the uncertainties of the
limb-darkening coefficients obtained using the confidencearea
method based on the $\chi^2_P$ distribution; the uncertainties are
given for $\gamma=0.955$.} \label{LD_x1_Xi2P_HD189733_lr}
\end{figure*}

Let us now consider the limb-darkening coefficients in the linear
limb-darkening law as a function of wavelength. Figures
\ref{LD_x1_diff_HD189733_left},
\ref{LD_x1_teor_obs_diff_HD189733_right},
\ref{LD_x1_diff_HD189733_lr}, \ref{LD_x1_Xi2P_HD189733_lr} show
the observed limbdarkening coefficients $x$ versus the wavelength
$\lambda$, together with the theoretical linear dependences of the
limb-darkening coefficients on ë in the photometric systems ugriz
and UBVRIJ \cite{Claret2004, Claret2000}. The observed relation
$x(\lambda)$ agrees qualitatively with the theoretical one: the
observed $x(\lambda)$ values decrease, on average, with increasing
wavelength. However, the observed $x(\lambda)$ are systematically
lower than the theoretical values. This is in qualitative
agreement with the results obtained for the HD 209458 system
\cite{Abubekerov2010, Southworth2008}.

However, the behavior of the observed $x(\lambda)$ differs from
the $x(\lambda)$ relation obtained for HD 209458, where the
difference between the observed and theoretical $x(\lambda)$
values increases monotonically with $\lambda$; for the HD 189733
system, the difference is largest at the shortest wavelengths and
decreases toward longer wavelengts. Figures
\ref{LD_x1_diff_HD189733_left} and
\ref{LD_x1_teor_obs_diff_HD189733_right} indicate that the results
obtained when fitting the left and right branches of the HD 189733
light curve separately agree with each other, despite the
systematic shift in brightness between these light curves.
Therefore, we conclude that our results for the limb-darkening of
the star are stable against systematic errors affecting the light
curve in the HD 189733 system.

\begin{table}[h!]
\caption{Joint fitting of the right and left branches of the
observed light curves for HD 189733 \cite{Pont2008} with the
quadratic limb-darkening law. (Parameter uncertainties estimated
using the differential-correction method are given at the
$2\sigma$ level. The two last columns present the values of the
reduced $\chi^2$ and the corresponding critical significance
levels $\alpha_0$).} \label{tabl_diff_Knutson_nonlin} \vspace{3mm}
\centering \tiny
\begin{tabular}{c|c|c|c|c|c|c|c|c|c|c|c|c}
$\lambda (\AA)$ & $r_s^c$ & $2\sigma
_{\text{est}}\left(r_s^c\right)$ & $r_p^c$ & $2\sigma
_{\text{est}}\left(r_p^c\right)$ & $i^c ({ }^{\circ})$ &
$2\sigma _{\text{est}}\left(i^c\right) ({ }^{\circ})$ & $x_1^c$ & $2\sigma _{\text{est}}\left(x_1^c\right)$ & $y_1^c$ & $2\sigma _{\text{est}}\left(y_1^c\right)$ & $\chi^2_{red}$ & $\alpha_0$\\
\hline
5750 & 0.11210 & 0.00067 & 0.01760 & 0.00015 & 85.705 & 0.058 & 0.57 & 0.18 & -0.02 & 0.27 & 5.5682 & 0 \\
6250 & 0.11228 & 0.00094 & 0.01792 & 0.00017 & 85.637 & 0.072 & 0.85 & 0.33 & -0.36 & 0.45 & 0.99780 & 0.608\\
6750 & 0.11183 & 0.00086 & 0.01768 & 0.00018 & 85.697 & 0.072 & 0.72 & 0.26 & -0.15 & 0.37 & 1.5358 & 0.00013 \\
7250 & 0.11156 & 0.00071 & 0.01756 & 0.00015 & 85.737 & 0.061 & 0.66 & 0.20 & -0.10 & 0.29 & 1.9376 & $8\cdot10^{-10}$ \\
7750 & 0.11180 & 0.00072 & 0.01759 & 0.00015 & 85.716 & 0.061 & 0.63 & 0.21 & -0.09 & 0.30 & 2.0992 & $2.5\cdot10^{-12}$ \\
8250 & 0.11201 & 0.00069 & 0.01764 & 0.00015 & 85.712 & 0.058 & 0.61 & 0.20 & -0.10 & 0.29 & 2.2615 & $5\cdot10^{-15}$\\
8750 & 0.11216 & 0.00077 & 0.01761 & 0.00016 & 85.695 & 0.065 & 0.58 & 0.24 & -0.10 & 0.34 & 2.0344 & $2.7\cdot10^{-11}$ \\
9250 & 0.11174 & 0.00090 & 0.01743 & 0.00022 & 85.766 & 0.084 & 0.43 & 0.23 & 0.09 & 0.34 & 1.4995 &  $3\cdot10^{-4}$ \\
9750 & 0.11105 & 0.00091 & 0.01736 & 0.00021 & 85.804 & 0.083 & 0.47 & 0.25 & 0.01 & 0.37 & 1.3637 & 0.0063 \\
10250 & 0.1129 & 0.0023 & 0.01733 & 0.00073 & 85.74 & 0.18 & 0.00 & 0.35 & 0.72 & 0.38 & 0.77561 & 0.98971\\
\hline
\end{tabular}
\end{table}

HD 189733 and HD 209458 have qualitatively different functions
$x(\lambda)$, probably due to the fact that there were spots on
the surface of the star in the HD 189733 system at the observing
epoch \cite{Pont2007, Pont2008}. Therefore, additional
observations of HD 189733 in periods of lower activity would be of
great interest. Note that the observed $x(\lambda)$ values in both
HD 189733 and HD 209458 lie below the theoretical relations,
although the observed $x(\lambda)$ relations are qualitatively
different for them. This result is important for checking modern
models of thin stellar atmospheres.

\section*{\normalsize FITTING OF THE LIGHT CURVES
WITH QUADRATIC LIMB DARKENING}

The results of fitting the light curves using the quadratic
limb-darkening law are given in Tables
\ref{tabl_diff_Knutson_nonlin_left} -- \ref{tabl_XiP_nonlin}. The
data obtained using the differential-correction method are given
in Tables \ref{tabl_diff_Knutson_nonlin_left} --
\ref{tabl_diff_Knutson_nonlin}. Here, the central parameter values
$r_p$, $r_s$, $i$, $x_1$, $y_1$ together with their $2\sigma$
uncertainties ($\gamma = 95.5\%$) are shown. Tables
\ref{tabl_diff_Knutson_nonlin_left} --
\ref{tabl_diff_Knutson_nonlin} also contain the values of
$\chi^2_{red}$ and $\alpha_0$. Fitting of the left and right
branches of the light curves (Tables
\ref{tabl_diff_Knutson_nonlin_left} and
\ref{tabl_diff_Knutson_nonlin_right}) yields $\chi^2_{red} < 1$
and $\alpha_0 > 0.5$ for some wavelengths. This indicates that the
observed brightnesses may be correlated and systematic errors may
be present in the observational data. When we fit the light curve
as a whole (Table \ref{tabl_diff_Knutson_nonlin}), our model is
"bad" for most wavelengths, and is rejected at a very low
significance level. The light-curve fits at $\lambda = 6250 \AA$
and $\lambda = 10250\AA$ yield $\chi^2_{red} \leq 1$ and $\alpha_0
> 0.5$, suggesting a strong correlation of the observational points.
All these factors lead us to adopt $2\sigma_{est}$ rather than
$\sigma_{est}$ to estimate the uncertainties of the model
parameters for the quadratic limb-darkening law, as was done for
the linear limb-darkening model.

\begin{table}[h!]
\caption{Fitting of the left branches of the observed light curves
for HD 189733 \cite{Pont2008} with the quadratic limb-darkening
law. (Parameter uncertainties were estimated using the
confidence-area method based on the $\chi^2_P$ distribution, with
$\gamma=0.955$.)} \label{tabl_XiP_nonlin_left} \vspace{3mm}
\centering \small
\begin{tabular}{c|c|c|c|c|c|c|c|c|c|c}
$\lambda (\AA)$ & $r_s$ & $\Delta _P\left(r_s\right)$ & $r_p$ &
$\Delta _P\left(r_p\right)$ & $i$ & $\Delta
_P\left(i\right)$ & $x_1$ & $\Delta_P\left(x_1\right)$ & $y_1$ & $\Delta_P\left(y_1\right)$\\
\hline
5750 & 0.11212 & 0.00067 & 0.01754 & 0.00016 & 85.722${}^{\circ}$ & 0.061${}^{\circ}$ & 0.43 & 0.18 & 0.17 & 0.27 \\
6750 & 0.1123 & 0.0020 & 0.01757 & 0.00047 & 85.71${}^{\circ}$ & 0.18${}^{\circ}$ & 0.51 & 0.48 & 0.13 & 0.73 \\
7250 & 0.1116 & 0.0013 & 0.01747 & 0.00031 & 85.76${}^{\circ}$ & 0.12${}^{\circ}$ & 0.51 & 0.32 & 0.07 & 0.49 \\
7750 & 0.1120 & 0.0013 & 0.01752 & 0.00028 & 85.72${}^{\circ}$ & 0.11${}^{\circ}$ & 0.45 & 0.32 & 0.08 & 0.48 \\
8250 & 0.1126 & 0.0012 & 0.01754 & 0.00029 & 85.72${}^{\circ}$ & 0.12${}^{\circ}$ & 0.32 & 0.34 & 0.29 & 0.51 \\
8750 & 0.1125 & 0.0014 & 0.01758 & 0.00030 & 85.69${}^{\circ}$ & 0.12${}^{\circ}$ & 0.41 & 0.37 & 0.09 & 0.54 \\
9250 & 0.1122 & 0.0015 & 0.01745 & 0.00038 & 85.76${}^{\circ}$ & 0.15${}^{\circ}$ & 0.32 & 0.45 & 0.20 & 0.67 \\
9750 & 0.1112 & 0.0018 & 0.01736 & 0.00039 & 85.80${}^{\circ}$ & 0.16${}^{\circ}$ & 0.43 & 0.39 & 0.13 & 0.62 \\
\hline
\end{tabular}
\end{table}

\begin{table}[h!]
\caption{Fitting of the left branches of the observed light curves
for HD 189733 \cite{Pont2008} with the quadratic limb-darkening
law. (Parameter uncertainties were estimated using the
confidence-area method based on the $\chi^2_M$ distribution, with
$\gamma=0.955$).} \label{tabl_XiM_nonlin_left} \vspace{3mm}
\centering \small
\begin{tabular}{c|c|c|c|c|c|c|c|c|c|c}
$\lambda (\AA)$ & $r_s$ & $\Delta _M\left(r_s\right)$ & $r_p$ &
$\Delta _M\left(r_p\right)$ & $i$ & $\Delta
_M\left(i\right)$ & $x_1$ & $\Delta_M\left(x_1\right)$ & $y_1$ & $\Delta_M\left(y_1\right)$\\
\hline
 6750 & 0.1150 & 0.0066 & 0.01785 & 0.00051 & 85.77${}^{\circ}$ & 0.41${}^{\circ}$ & 1.27 & 0.81 & 1.3 & 1.3 \\
 7250 & 0.1115 & 0.0022 & 0.01758 & 0.00039 & 85.77${}^{\circ}$ & 0.21${}^{\circ}$ & 0.63 & 0.40 & 0.23 & 0.62 \\
 7750 & 0.1118 & 0.0014 & 0.01756 & 0.00030 & 85.71${}^{\circ}$ & 0.12${}^{\circ}$ & 0.47 & 0.27 & 0.13 & 0.42 \\
 8750 & 0.1123 & 0.0018 & 0.01766 & 0.00036 & 85.69${}^{\circ}$ & 0.17${}^{\circ}$ & 0.51 & 0.36 & 0.28 & 0.55 \\
 9250 & 0.1121 & 0.0022 & 0.01762 & 0.00036 & 85.76${}^{\circ}$ & 0.22${}^{\circ}$ & 0.42 & 0.35 & 0.36 & 0.56 \\
 9750 & 0.1109 & 0.0018 & 0.01741 & 0.00040 & 85.77${}^{\circ}$ & 0.14${}^{\circ}$ & 0.39 & 0.40 & 0.073 & 0.64 \\
\hline
\end{tabular}
\end{table}

\begin{table}[h!]
\caption{Fitting of the right branches of the observed light
curves for HD 189733 \cite{Pont2008} with the quadratic
limb-darkening law. (Parameter uncertainties were estimated using
the confidence-area method based on the $\chi^2_P$ distribution,
with $\gamma=0.955$.)} \label{tabl_XiP_nonlin_right} \vspace{3mm}
\centering \small
\begin{tabular}{c|c|c|c|c|c|c|c|c|c|c}
$\lambda (\AA)$ & $r_s$ & $\Delta _P\left(r_s\right)$ & $r_p$ &
$\Delta _P\left(r_p\right)$ & $i$ & $\Delta
_P\left(i\right)$ & $x_1$ & $\Delta_P\left(x_1\right)$ & $y_1$ & $\Delta_P\left(y_1\right)$\\
\hline
5750 & 0.11091 & 0.00070 & 0.01752 & 0.00016 & 85.782${}^{\circ}$ & 0.060${}^{\circ}$ & 0.87 & 0.18 & -0.44 & 0.26 \\
6250 & 0.1109 & 0.0028 & 0.01784 & 0.00064 & 85.73${}^{\circ}$ & 0.22${}^{\circ}$ & 1.27 & 0.69 & -0.73 & 0.97 \\
6750 & 0.1108 & 0.0019 & 0.01758 & 0.00043 & 85.77${}^{\circ}$ & 0.16${}^{\circ}$ & 1.02 & 0.44 & -0.60 & 0.64 \\
7250 & 0.1111 & 0.0014 & 0.01750 & 0.00037 & 85.78${}^{\circ}$ & 0.13${}^{\circ}$ & 0.85 & 0.37 & -0.39 & 0.54 \\
7750 & 0.1111 & 0.0013 & 0.01751 & 0.00030 & 85.78${}^{\circ}$ & 0.11${}^{\circ}$ & 0.87 & 0.33 & -0.38 & 0.48 \\
8250 & 0.1108 & 0.0012 & 0.01763 & 0.00027 & 85.778${}^{\circ}$ & 0.098${}^{\circ}$ & 1.01 & 0.29 & -0.59 & 0.42 \\
8750 & 0.1113 & 0.0015 & 0.01753 & 0.00032 & 85.76${}^{\circ}$ & 0.12${}^{\circ}$ & 0.85 & 0.38 & -0.45 & 0.55 \\
9250 & 0.1109 & 0.0016 & 0.01731 & 0.00041 & 85.84${}^{\circ}$ & 0.16${}^{\circ}$ & 0.66 & 0.45 & -0.12 & 0.67 \\
9750 & 0.1108 & 0.0020 & 0.01726 & 0.00043 & 85.85${}^{\circ}$ & 0.18${}^{\circ}$ & 0.55 & 0.53 & -0.17 & 0.78 \\
\hline
\end{tabular}
\end{table}

\begin{table}[h!]
\caption{Joint fitting of the right and left branches of the
observed light curves for HD189733 \cite{Pont2008} with the
quadratic limbdarkening law. (Parameter uncertainties were
estimated using the confidence-area method based on the $\chi^2_P$
distribution, with $\gamma=0.955$.)} \label{tabl_XiP_nonlin}
\vspace{3mm} \centering \small
\begin{tabular}{c|c|c|c|c|c|c|c|c|c|c}
$\lambda (\AA)$ & $r_s$ & $\Delta _P\left(r_s\right)$ & $r_p$ &
$\Delta _P\left(r_p\right)$ & $i$ & $\Delta
_P\left(i\right)$ & $x_1$ & $\Delta_P\left(x_1\right)$ & $y_1$ & $\Delta_P\left(y_1\right)$\\
\hline
5750 & 0.11182 & 0.00048 & 0.01757 & 0.00011 & 85.724${}^{\circ}$ & 0.042${}^{\circ}$ & 0.60 & 0.12 & -0.09 & 0.18 \\
6250 & 0.1124 & 0.0021 & 0.01790 & 0.00047 & 85.64${}^{\circ}$ & 0.17${}^{\circ}$ & 0.88 & 0.53 & -0.35 & 0.76 \\
6750 & 0.1119 & 0.0014 & 0.01765 & 0.00031 & 85.71${}^{\circ}$ & 0.12${}^{\circ}$ & 0.71 & 0.34 & -0.16 & 0.50 \\
7250 & 0.11160 & 0.00095 & 0.01755 & 0.00023 & 85.741${}^{\circ}$ & 0.085${}^{\circ}$ & 0.65 & 0.24 & -0.12 & 0.36 \\
7750 & 0.11183 & 0.00091 & 0.01758 & 0.00019 & 85.718${}^{\circ}$ & 0.075${}^{\circ}$ & 0.62 & 0.23 & -0.10 & 0.34 \\
8250 & 0.11204 & 0.00085 & 0.01763 & 0.00019 & 85.714${}^{\circ}$ & 0.072${}^{\circ}$ & 0.61 & 0.22 & -0.11 & 0.32 \\
8750 & 0.11219 & 0.00100 & 0.01760 & 0.00021 & 85.697${}^{\circ}$ & 0.082${}^{\circ}$ & 0.58 & 0.27 & -0.10 & 0.38 \\
9250 & 0.1118 & 0.0011 & 0.01743 & 0.00028 & 85.77${}^{\circ}$ & 0.11 & 0.45 & 0.33 & 0.12 & 0.48 \\
9750 & 0.1111 & 0.0013 & 0.01735 & 0.00027 & 85.81${}^{\circ}$ & 0.11 & 0.48 & 0.35 & 0.01 & 0.51 \\
\hline
\end{tabular}
\end{table}

Tables \ref{tabl_XiP_nonlin_left}, \ref{tabl_XiP_nonlin_right},
\ref{tabl_XiP_nonlin} show the results of fitting the light curves
of HD 189733 using the model with quadratic limb darkening; the
parameter uncertainties are estimated using the confidence-area
method based on the $\chi^2_P$ distribution. Table
\ref{tabl_XiM_nonlin_left} presents the results of fitting the
left branches of the light curves using the quadratic
limb-darkening law. Here, the parameter uncertainties were
obtained using the confidence-area method based on the $\chi^2_M$
distribution ($\gamma = 0.955$).

\section*{\normalsize $x_1$ AND $y_1$ AS FUNCTIONS
OF WAVELENGTH IN THE QUADRATIC LIMB-DARKENING LAW}

\begin{figure*}[h!]
\vspace{0cm} \epsfxsize=0.99\textwidth
\epsfbox{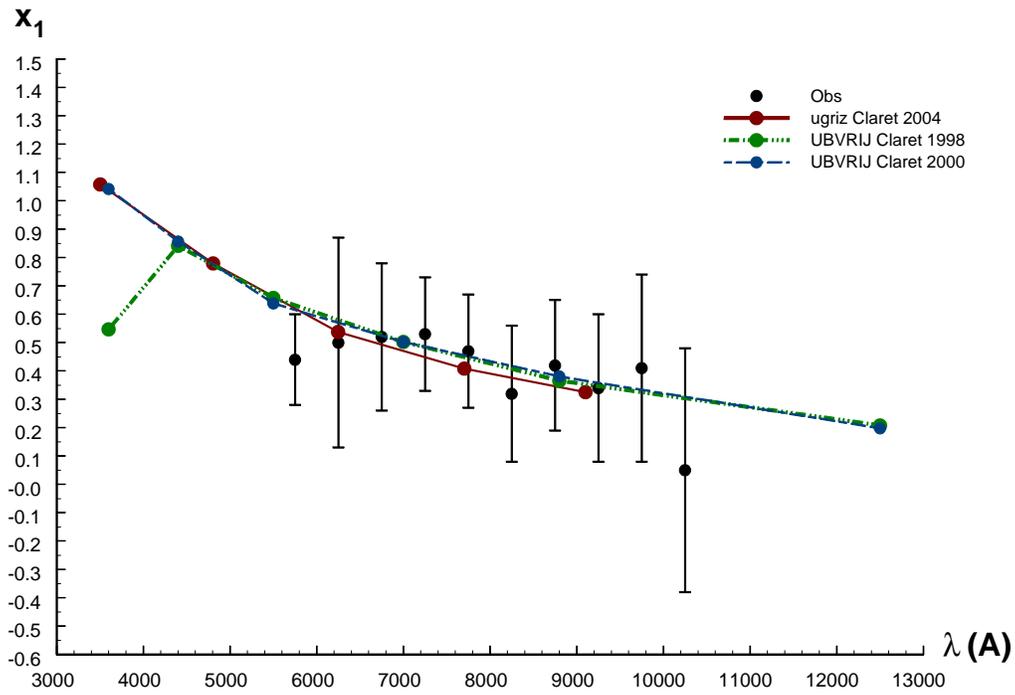}
\caption{Same function as in Fig.4 but for the linear coefficient
$x_1$ and the quadratic limb-darkening law.}
\label{x1_quadr_err_diffpopr_HD189_left}
\end{figure*}

\begin{figure*}[h!]
\vspace{0cm} \epsfxsize=0.99\textwidth
\epsfbox{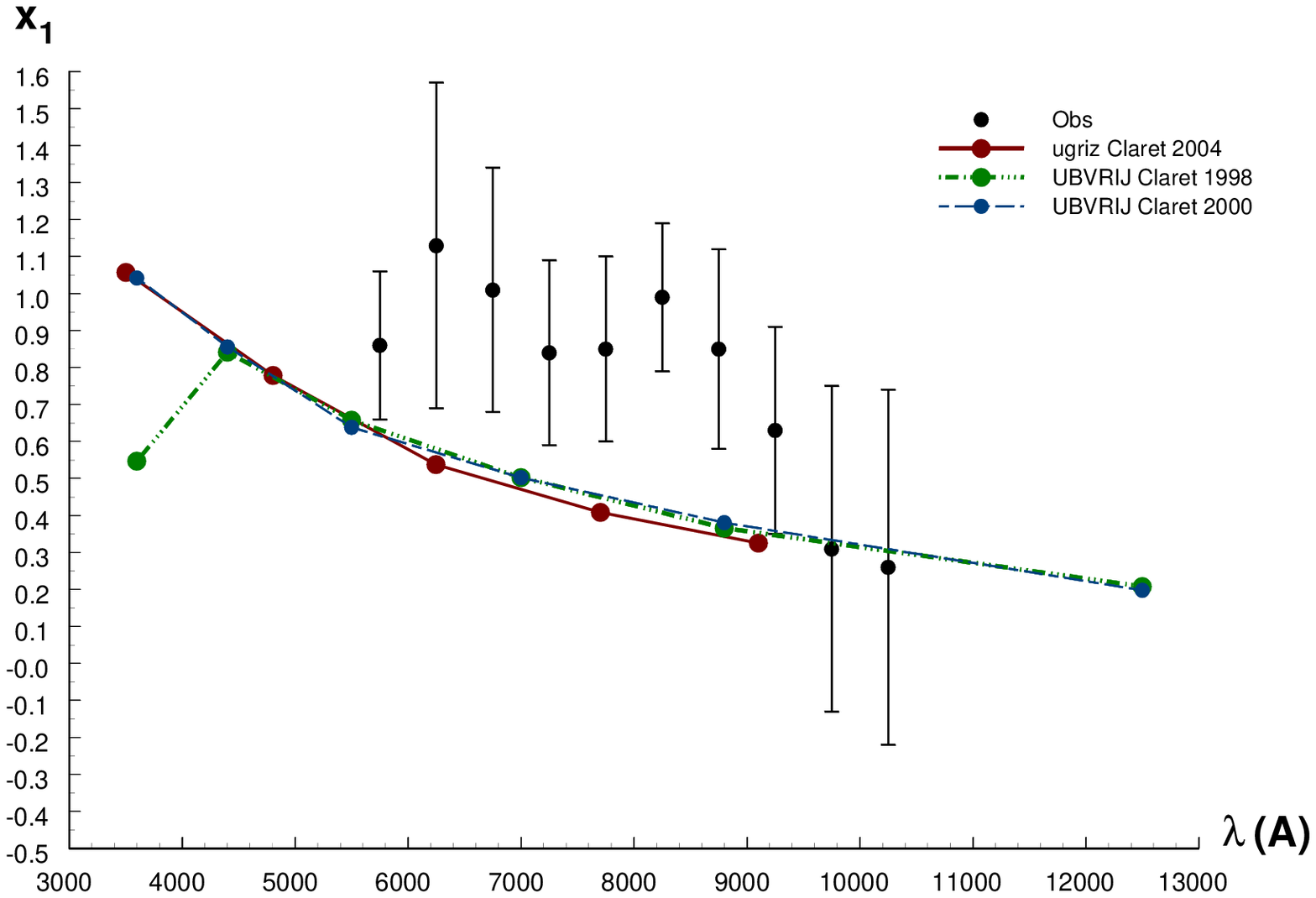}
\caption{Same function as in Fig. 8 but obtained from the analysis
of the right branches of the light curves.}
\label{x1_quadr_err_diffpopr_HD189_right}
\end{figure*}

\begin{figure*}[h!]
\vspace{0cm} \epsfxsize=0.99\textwidth
\epsfbox{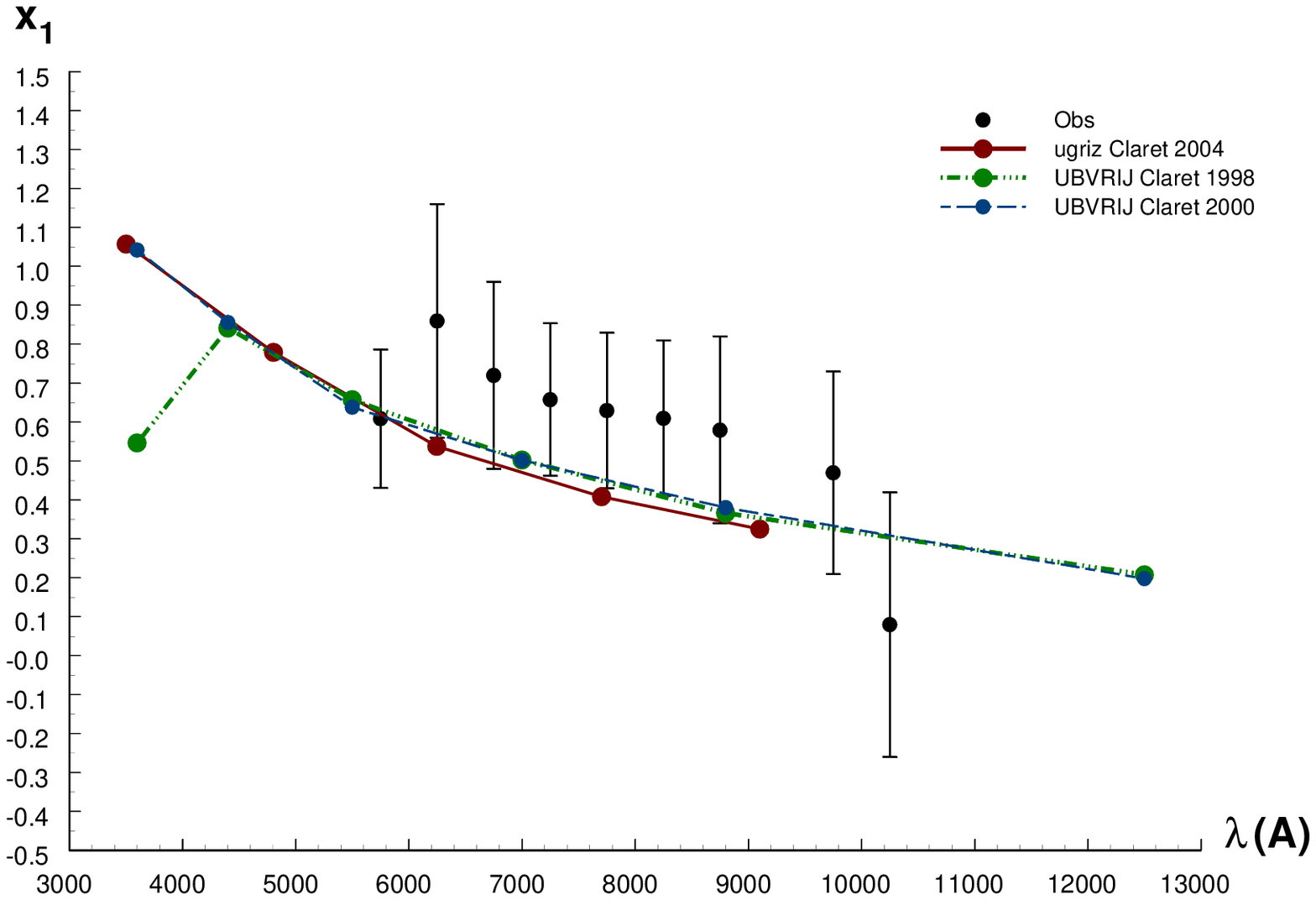} \caption{Same
function as in Fig. 8 but obtained from the joint analysis of the
left and right branches of the light curves.}
\label{x1_quadr_err_diffpopr_HD189_lr}
\end{figure*}

\begin{figure*}[h!]
\vspace{0cm} \epsfxsize=0.99\textwidth
\epsfbox{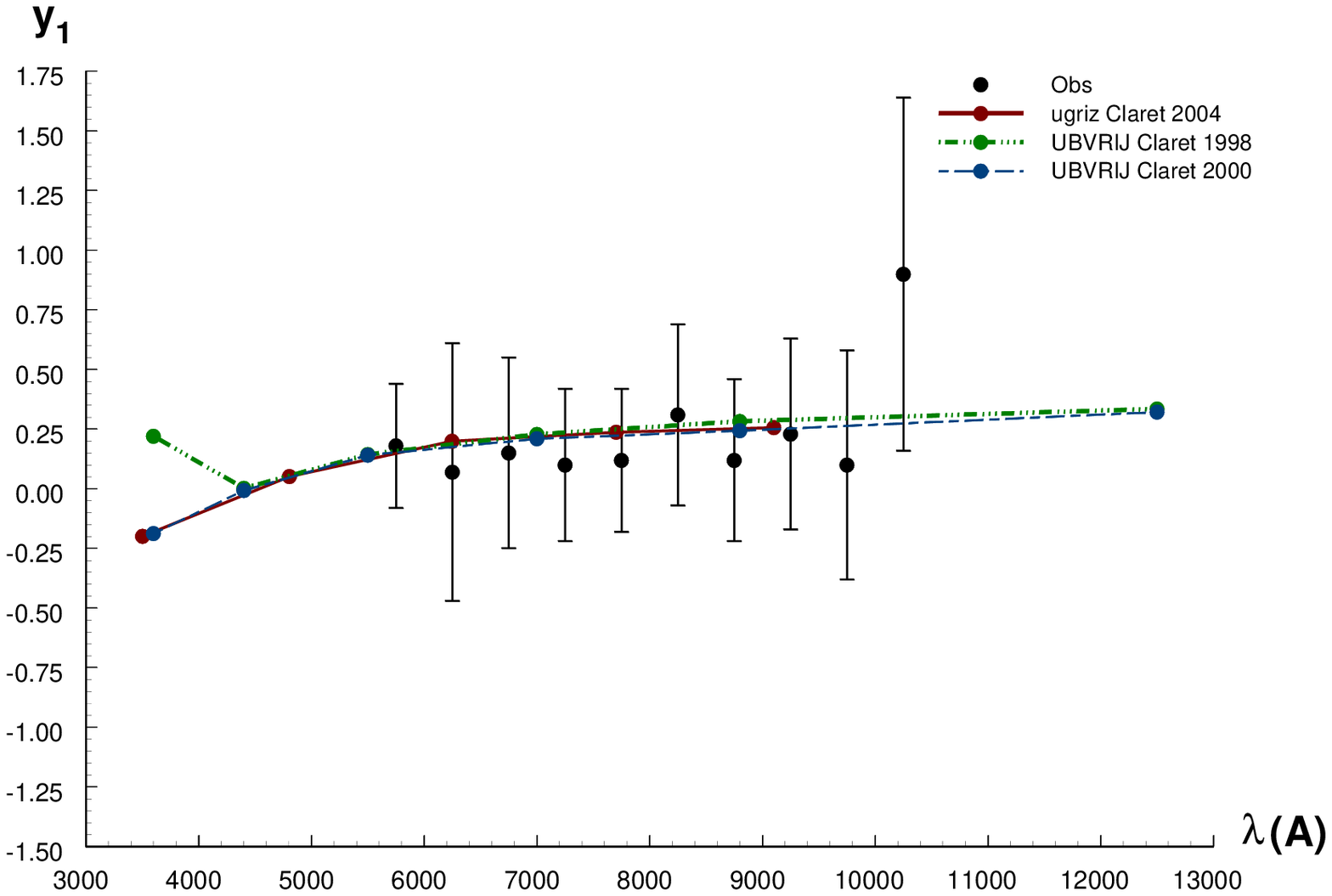}
\caption{Same function as in Fig.8 but for the quadratic
coefficient $y_1$.} \label{y1_quadr_err_diffpopr_HD189_left}
\end{figure*}

\begin{figure*}[h!]
\vspace{0cm} \epsfxsize=0.99\textwidth
\epsfbox{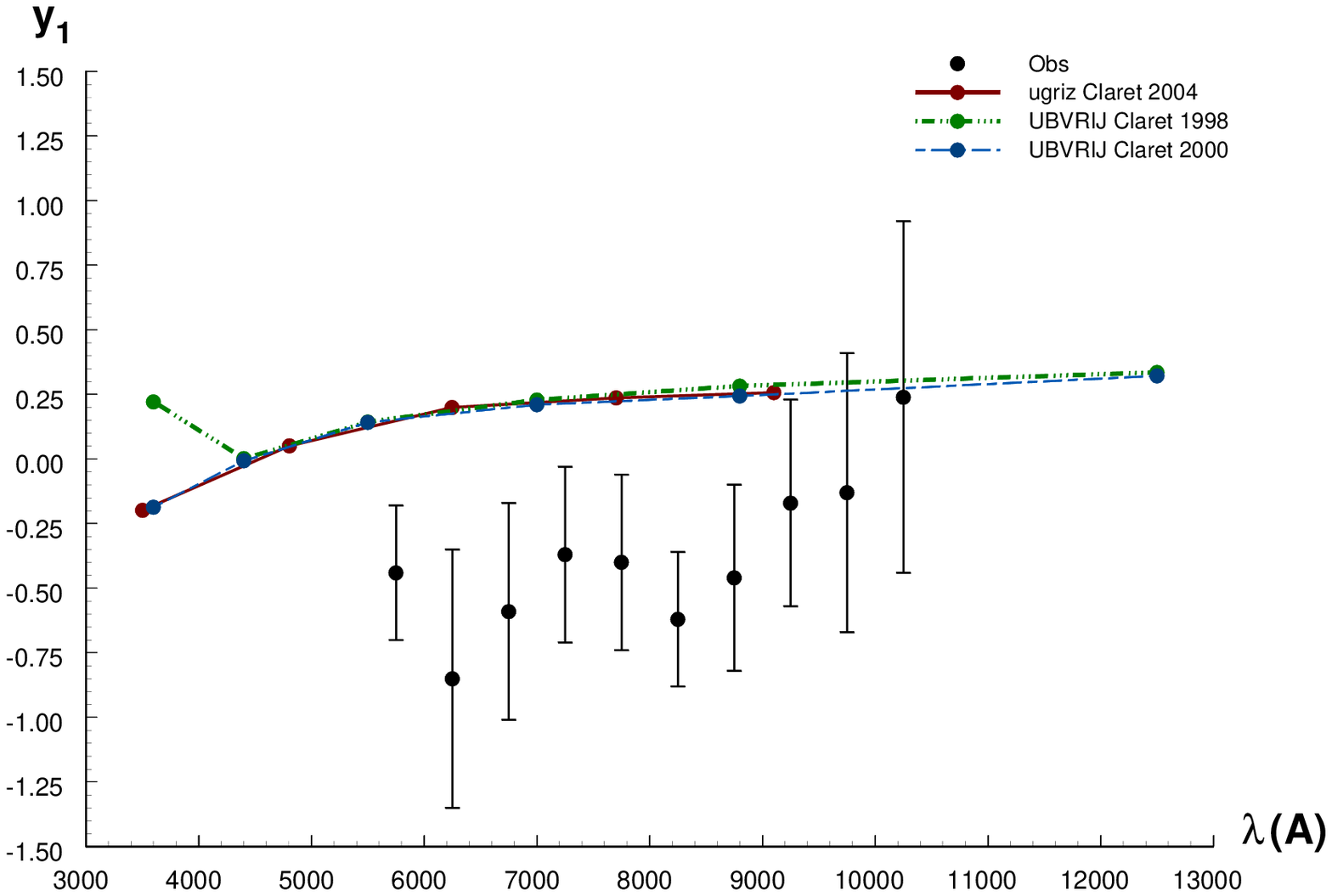}
\caption{Same function as in Fig.11 but obtained for the analysis
of the right branches of the light curves.}
\label{y1_quadr_err_diffpopr_HD189_right}
\end{figure*}

\begin{figure*}[h!]
\vspace{0cm} \epsfxsize=0.99\textwidth
\epsfbox{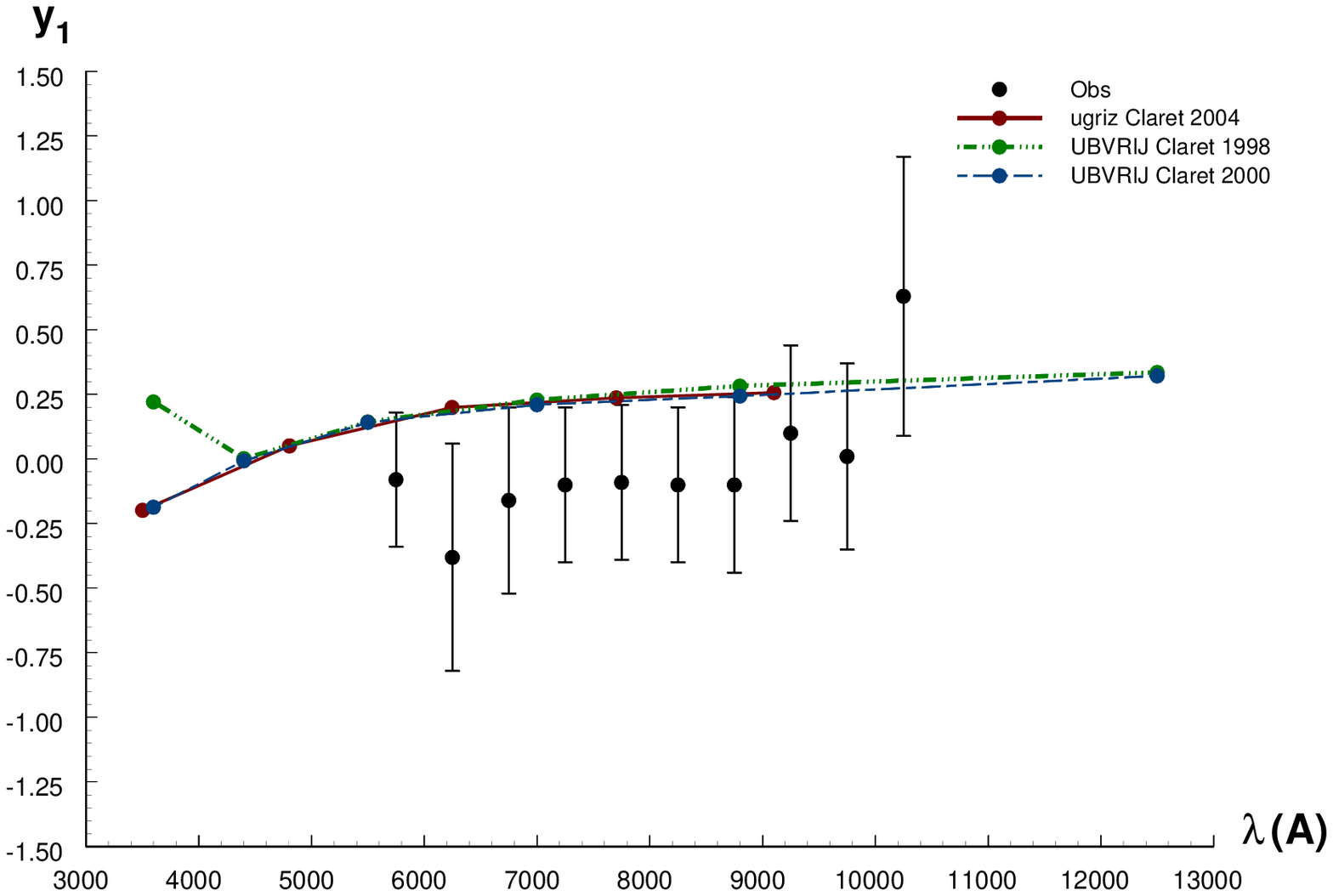} \caption{Same
function as in Fig. 11 but obtained for the joint analysis of the
left and right branches of the light curves.}
\label{y1_quadr_err_diffpopr_HD189_lr}
\end{figure*}

Figures \ref{x1_quadr_err_diffpopr_HD189_left} --
\ref{y1_quadr_err_diffpopr_HD189_lr} show the observed
$x_1(\lambda)$ and $y_1(\lambda)$ values as functions of
wavelength, obtained using the differential-correction method. The
$2\sigma$ uncertainties are given ($\gamma = 95.5\%$). The
theoretical relations $x_1(\lambda)$ and $y_1(\lambda)$ obtained
in \cite{Claret2004,Claret2000} are also shown here. The observed
relation $x_1(\lambda)$ obtained using the left branch of the
light curve agrees with the theoretical function $x_1(\lambda)$,
while the observed $x_1(\lambda)$ obtained using the right branch
of the light curve lies considerably above the theoretical
relation (Figs.\ref{x1_quadr_err_diffpopr_HD189_left} and
\ref{x1_quadr_err_diffpopr_HD189_right}). The observed
$x_1(\lambda)$ obtained using the whole light curve agrees
satisfactorily with the theoretical relation within the
uncertainties (at the $2\sigma$ level).

The observed $y_1(\lambda)$ obtained using the left branch of the
light curve agrees with the theoretical function, while the
observed $y_1(\lambda)$ obtained for the right branch of the light
curve lies considerably below the theoretical relation. The
observed $y_1(\lambda)$ obtained using the whole light curve
agrees satisfactorily with the theoretical relation within the
uncertainties (at the $2\sigma$ level) (Figs.
\ref{y1_quadr_err_diffpopr_HD189_left} -
\ref{y1_quadr_err_diffpopr_HD189_lr}).

\begin{figure*}[h!]
\vspace{0cm} \epsfxsize=0.99\textwidth
\epsfbox{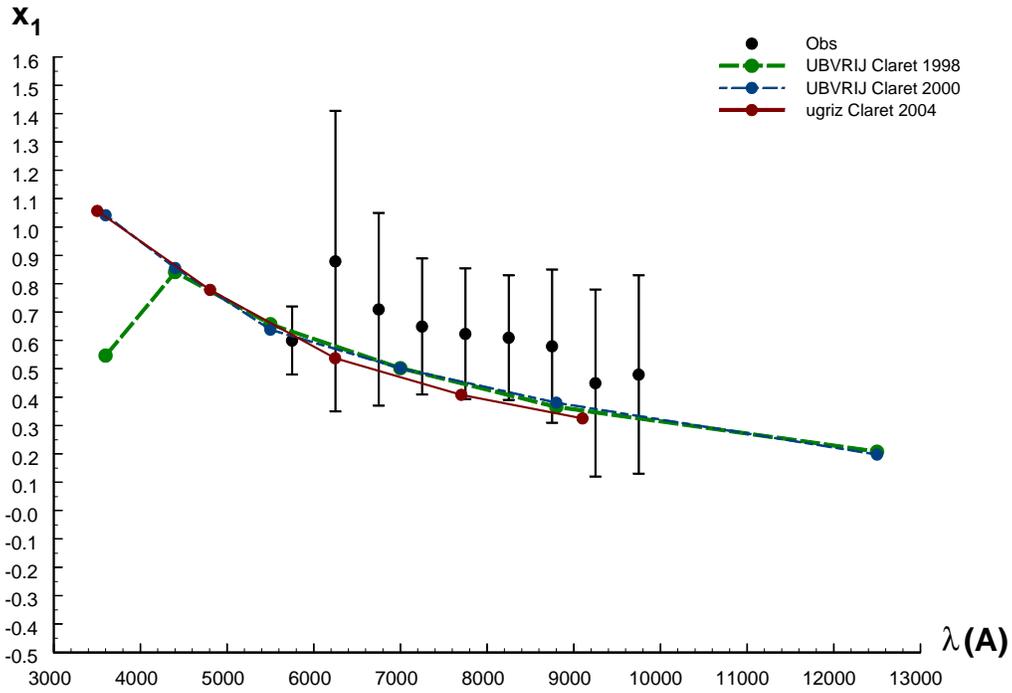} \caption{Same
function as in Fig.10 but with the uncertainties in limb-darkening
coefficients calculated using the confidencearea method based on
the $\chi^2_P$ distribution; uncertainties correspond to
$\gamma=0.955$.} \label{x1_quadr_err_Xi2P_HD189_lr}
\end{figure*}

\begin{figure*}[h!]
\vspace{0cm} \epsfxsize=0.99\textwidth
\epsfbox{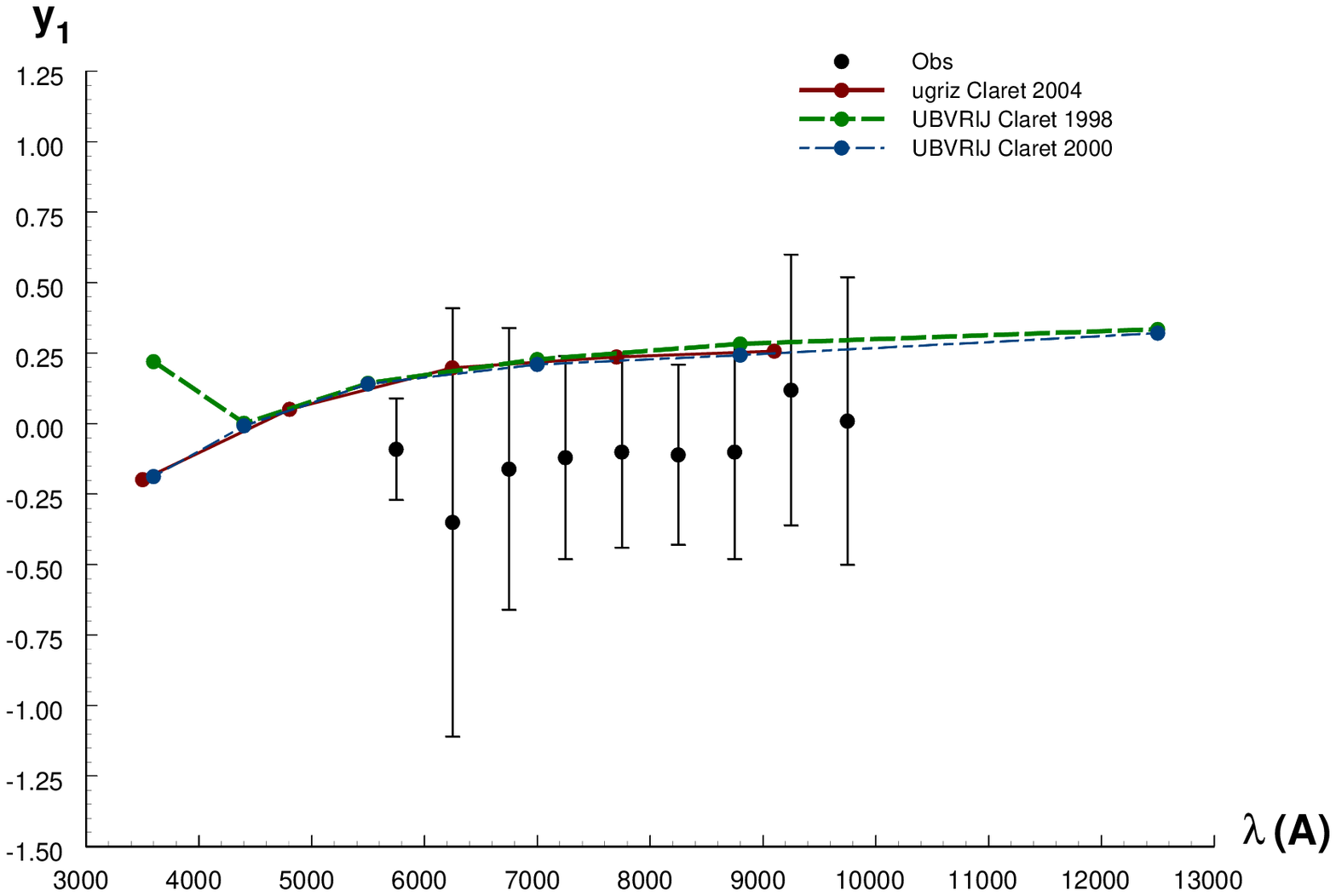} \caption{Same
function as in Fig. 14 but for the quadratic coefficient $y_1$.}
\label{y1_quadr_err_Xi2P_HD189_lr}
\end{figure*}

Note that the observed relations $x_1(\lambda)$ and $y_1(\lambda)$
obtained using the whole light curve do not agree with the
theoretical relations at the $1\sigma$ level: the observed
$x_1(\lambda)$ is systematically higher than the theoretical
relation, and $y_1(\lambda)$ is lower than the theoretical
relation (Figs. \ref{x1_quadr_err_diffpopr_HD189_lr} and
\ref{y1_quadr_err_diffpopr_HD189_lr}). Since our model is formally
"bad" and, moreover, has five rather than one parameters, we are
forced to take the uncertainties at the $2\sigma$ level. With
these uncertainties for the coefficients $x_1(\lambda)$ and
$y_1(\lambda)$, the differences between the observed and
theoretical $x_1(\lambda)$ and $y_1(\lambda)$ relations do not
seem to be significant. Figures \ref{x1_quadr_err_Xi2P_HD189_lr}
and \ref{y1_quadr_err_Xi2P_HD189_lr} show the $x_1(\lambda)$ and
$y_1(\lambda)$ relations obtained when fitting the left and right
branches of the light curve, with the uncertainties derived using
the confidence-area method based on the $\chi^2_P$ distribution.

\begin{figure*}[h!]
\vspace{0cm} \epsfxsize=0.99\textwidth
\epsfbox{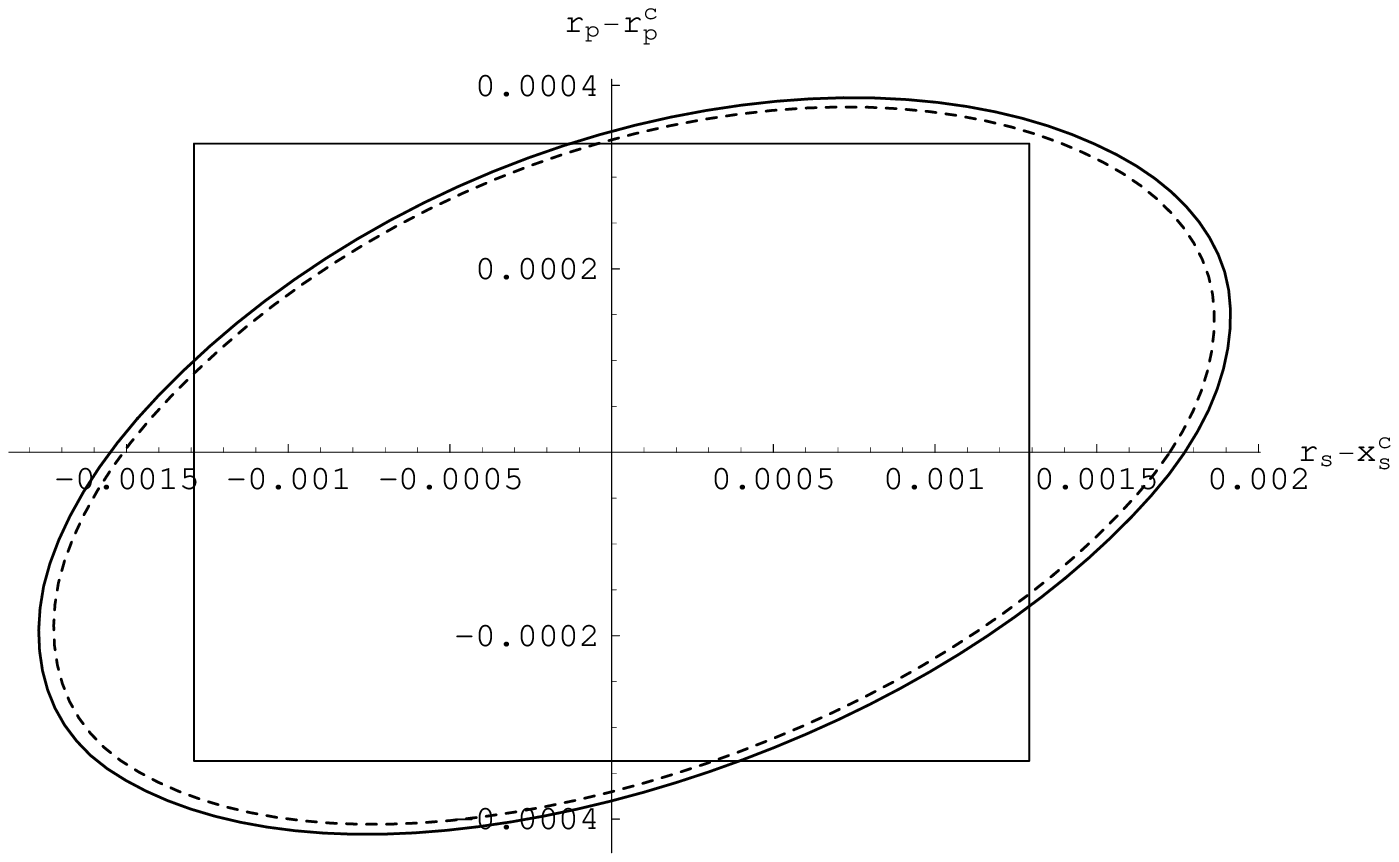} \caption{Projections of the
confidence area (for $\gamma=0.955$) obtained using the $\chi^2_P$
(dashed line) and $\chi^2_M$ (solid line) distributions in the
plane of the parameters $r_s, r_p$ when fitting the left branch of
the light curve for $9500 - 10000 \AA$ with the quadratic
limb-darkening law. The sides of the rectangle correspond to the
$2\sigma$ error intervals obtained using the
differential-correction method.} \label{Section9rsrp2XiP95}
\end{figure*}

\begin{figure*}[h!]
\vspace{0cm} \epsfxsize=0.99\textwidth
\epsfbox{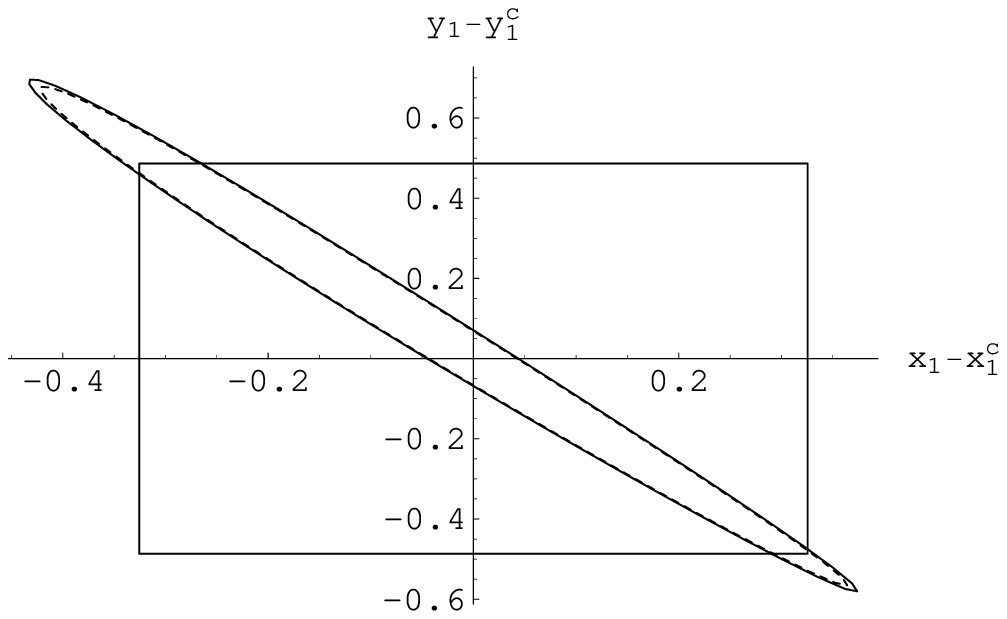} \caption{Projections of the
confidence area (for $\gamma=0.955$) obtained using the $\chi^2_P$
(dashed line) and $\chi^2_M$ (solid line) distributions in the
plane of the parameters $x_1, y_1$ when fitting the left branch of
the light curve for $9500 - 10000 \AA$ with the quadratic
limb-darkening law. The sides of the rectangle correspond to the
$2\sigma$ error intervals obtained using the
differential-correction method.} \label{Section9x1y1XiP95}
\end{figure*}

Figures \ref{Section9rsrp2XiP95} and \ref{Section9x1y1XiP95}
illustrate the projections of the confidence areas in the ($r_s,
r_p$) and ($x_1, y_1$) planes obtained when using the $\chi_P^2$
and $\chi_M^2$ distributions. The light curve at $\lambda 9500\div
10000 \AA$ (left branch) was used with the model based on the
quadratic limbdarkening law.

\section*{\normalsize CONCLUSION}

We have analyzed high-accuracy multicolor light curves given in
\cite{Pont2007, Pont2008} for the star–explanet system HD 189733
and determined the radii of the star and planet, the orbital
inclination, and the coefficients in the linear and quadratic
limb-darkening laws across the disk of the K2V star. The results
of our fitting agree with the data obtained in \cite{Pont2007,
Pont2008}.

We have analyzed in detail the limb darkening across the disk of
the K2V star based on the resulting uncertainties in the
coefficients in the linear and quadratic limb-darkening laws. We
allowed for the presence of spots on the surface of the K2V star
\cite{Pont2008} by analyzing the left and right branches of the
light curves separately, as well as the light curve as a whole.
Moreover, we paid special attention to checking the adequacy of
our model and verifying how well it agrees with the observational
data. Our model proved to be formally "bad". When fitting the left
and right branches of the light curve separately, our model is
rejected at a high significance level of $\alpha_0
> 50\%$ at some wavelengths; this most likely indicates a correlation of the individual
observational points in the light curve. When fitting the light
curve as a whole, our model is rejected at a very low significance
level for most wavelengths.

Since our model turned out to be "bad" when applied to HD 189733,
we were forced to adopt the high confidence level $\gamma =
95.5\%$ (rather than the 68\% relevant for "good" models) when
estimating the parameter uncertainties. Since our model includes
four or five parameters instead of only one, we choose the error
intervals in the confidence-area method such that the probability
for them to encompass the true parameter values is undoubtedly
higher than the given probability (in this case, the given
confidence level $\gamma = 95.5\%$ is associated with the entire
confidence area $D$ rather than only one confidence interval). Our
analysis of the relations $x(\lambda)$, $x_1(\lambda)$,
$y_1(\lambda)$ for the limb-darkening coefficients deduced from
fitting the observations yielded the following results $\gamma =
95.5\%$.

The observed values of the coefficient $x(\lambda)$ in the linear
limb-darkening law for HD 189733 are systematically below the
theoretical relation, with the differences increasing with
decreasing wavelength $\lambda$ (in contrast to HD 209458, where
they increase with increasing $\lambda$). The observational
coefficients in the quadratic limb-darkening law $x_1(\lambda)$
and $y_1(\lambda)$ agree satisfactorily within the $2\sigma$
uncertainties ($\gamma = 95.5\%$) with the theoretical relations
developed for one-dimensional thin stellar atmospheres
(\cite{Claret2004, Claret2009, Claret2000}).

We emphasize that these conclusions concern the light curves of
the HD 189733 system when there were spots on the disk of the K2V
star. To further invetigate limb darkening on this star,
additional observational data should be obtained for eclipse
curves at epochs when the contribution of spots is negligible.

We confirm the earlier conclusion \cite{Pont2007} (at the
$1\sigma$ level) that the exoplanet radius increases with
decreasing wavelength. This may imply the presence of the
atmosphere around the exoplanet.

The authors are grateful to Frederic Pont for providing us with
the observational data on HD 189733.

\newpage

\newpage

\renewcommand{\figurename}{Fig}

\renewcommand{\figurename}{Fig}

\renewcommand{\figurename}{Fig}

\renewcommand{\figurename}{Fig}

\renewcommand{\figurename}{Fig}

\renewcommand{\figurename}{Fig}

\renewcommand{\figurename}{Fig}


\renewcommand{\figurename}{Fig}

\renewcommand{\figurename}{Fig}

\renewcommand{\figurename}{Fig}


\renewcommand{\figurename}{Fig}

\renewcommand{\figurename}{Fig}

\renewcommand{\figurename}{Fig}

\renewcommand{\figurename}{Fig}

\renewcommand{\figurename}{Fig}

\renewcommand{\figurename}{Fig}

\renewcommand{\figurename}{Fig}


\begin{thebibliography}{99}
\bibitem{Abubekerov2010}  M.K. Abubekerov, N.Yu. Gostev, and A.
M. Cherepashchuk, Astron. Rep. {\bf 54}, 1105 (2010).
\bibitem{Claret2004} A. Claret, Astron \& Astrophys {\bf 428},
1001 (2004).
\bibitem{Claret2009} A. Claret, Astron \& Astrophys {\bf 506},
1335 (2009).
\bibitem{Southworth2008} J. Southworth, Monthly Not. Roy. Astron.
Soc. {\bf 386}, 1644 (2008).
\bibitem{Pont2007} F.Pont, R.L. Gilliland, C. Moutou et al.,
Astron \& Astrophys {\bf 476}, 1347 (2007).
\bibitem{Pont2008} F. Pont, H. Knutson, R. L. Gilliland et al.,
Monthly Not. Roy. Astron. Soc. {\bf 385}, 109 (2008).
\bibitem{Bouchy2005} F. Bouchy, S. Udry, M. Mayor et al., Astron
\& Astrophys {\bf 444}, 15 (2005).
\bibitem{Kudzei1985} I. Kudzei, Astron. Tsirk., No. 1363 (1985).
\bibitem{Kasuya2011} S. Kasuya, M.Honda, R. Mishima, Monthly Not. Roy. Astron. Soc. {\bf 411}, 1863 (2011).
\bibitem{Abubekerov2008} M.K. Abubekerov, N.Yu. Gostev, and A.M.
Cherepashchuk, Astron. Rep. {\bf 52}, 99 (2008).
\bibitem{Abubekerov2009}  M.K. Abubekerov, N.Yu. Gostev, and A.M.
Cherepashchuk, Astron. Rep. {\bf 53}, 722 (2009).
\bibitem{Popper1984} D.M. Popper, Astron.J. {\bf 89}, 132 (1984).
\bibitem{Cherepashchuk1993} A.M. Cherepashchuk, Astron. Rep. {\bf 37}, 585 (1993).
\bibitem{Claret1998} A. Claret, Astron \& Astrophys {\bf 335}, 647 (1998).
\bibitem{Claret2000} A. Claret, Astron \& Astrophys {\bf 363}, 1081 (2000).
\bibitem{Eggleton1983} Eggleton P.P., Astrophys.J. {\bf 268}, 368 (1983).
\end{thebibliography}
\end{document}